\documentclass[longauth]{aa} % for the long lists of affiliations 
\usepackage{graphicx}
\usepackage{placeins}
\usepackage{txfonts}
\usepackage{hyperref}
\usepackage{orcidlink}
\newcommand{\orcit}[1]{\protect\href{https://orcid.org/#1}{\protect\includegraphics[width=8pt]{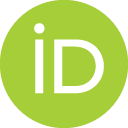}}}

\begin{document}

%   \title{BL Lacertae: the smoking gun of/clues to orientation changes in a curved jet}

   \title{A wiggling filamentary jet at the origin of the blazar multi-wavelength behaviour}

%   \subtitle{I. Overviewing the $\kappa$-mechanism}

\author{
       C. M. Raiteri\orcit{0000-0003-1784-2784}$^{ 1}$
\and           M. Villata\orcit{0000-0003-1743-6946}$^{ 1}$
\and      M. I. Carnerero\orcit{0000-0001-5843-5515}$^{ 1}$
\and     S. O. Kurtanidze$^{ 2}$
\and     D. O. Mirzaqulov\orcit{0000-0003-0570-6531}$^{ 3}$
\and         E. Ben\'itez\orcit{0000-0003-1018-2613}$^{ 4}$
\and           G. Bonnoli\orcit{0000-0003-2464-9077}$^{ 5}$
\and          D. Carosati\orcit{0000-0001-5252-1068}$^{ 6, 7}$
\and  J. A. Acosta-Pulido\orcit{0000-0002-0433-9656}$^{ 8}$
\and             I. Agudo\orcit{0000-0002-3777-6182}$^{ 9}$
\and       T. S. Andreeva\orcit{0000-0003-3613-6252}$^{10,11}$
\and          G. Apolonio$^{12}$
\and            R. Bachev\orcit{0000-0002-0766-864X}$^{13}$
\and         G. A. Borman\orcit{0000-0002-7262-6710}$^{14}$
\and          V. Bozhilov\orcit{0000-0002-3117-7197}$^{15}$
\and          L. F. Brown\orcit{0009-0000-9424-5575}$^{16}$
\and         W. Carbonell\orcit{0009-0008-0028-1670}$^{16}$
\and           C. Casadio\orcit{0000-0003-1117-2863}$^{17}$
\and           W. P. Chen\orcit{0000-0003-0262-272X}$^{18}$
\and       G. Damljanovic\orcit{0000-0002-6710-6868}$^{19}$
\and   S. A. Ehgamberdiev\orcit{0000-0001-9730-3769}$^{ 3,20}$
\and         D. Elsaesser\orcit{0000-0001-6796-3205}$^{21,22}$
\and          J. Escudero\orcit{0000-0002-4131-655X}$^{ 9}$
\and             M. Feige$^{21}$
\and           A. Fuentes\orcit{0000-0002-8773-4933}$^{ 9}$
\and         D. Gabellini$^{23}$
\and            K. Gazeas\orcit{0000-0002-8855-3923}$^{24}$
\and         M. Giroletti\orcit{0000-0002-8657-8852}$^{25}$
\and       T. S. Grishina\orcit{0000-0002-3953-6676}$^{26}$
\and          A. C. Gupta\orcit{0000-0002-9331-4388}$^{27,28}$
\and        M. A. Gurwell\orcit{0000-0003-0685-3621}$^{29}$
\and    V. A. Hagen-Thorn\orcit{0000-0002-6431-8590}$^{26}$
\and          G. M. Hamed\orcit{0000-0001-6009-1897}$^{30}$
\and           D. Hiriart\orcit{0000-0002-4711-7658}$^{31}$
\and            M. Hodges$^{32}$
\and       R. Z. Ivanidze$^{ 2}$
\and         D. V. Ivanov\orcit{0000-0002-9270-5926}$^{10,11}$
\and          M. D. Joner\orcit{0000-0003-0634-8449}$^{12}$
\and        S. G. Jorstad\orcit{0000-0001-6158-1708}$^{33,26}$
\and      M. D. Jovanovic\orcit{0000-0003-4298-3247}$^{19}$
\and         S. Kiehlmann$^{17,34}$
\and      G. N. Kimeridze$^{ 2}$
\and     E. N. Kopatskaya\orcit{0000-0001-9518-337X}$^{26}$
\and       Yu. A. Kovalev\orcit{0000-0002-8017-5665}$^{35,36}$
\and        Y. Y. Kovalev\orcit{0000-0001-9303-3263}$^{37}$
\and     O. M. Kurtanidze$^{ 2,38}$
\and         A. Kurtenkov$^{13}$
\and      E. G. Larionova\orcit{0000-0002-2471-6500}$^{26}$
\and           A. Lessing$^{21}$
\and            H. C. Lin$^{18}$
\and        J. M. L\'opez$^{39}$
\and             C. Lorey$^{21}$
\and            J. Ludwig$^{21}$
\and          N. Marchili\orcit{0000-0002-5523-7588}$^{25}$
\and          A. Marchini\orcit{0000-0003-3779-6762}$^{40}$
\and       A. P. Marscher\orcit{0000-0001-7396-3332}$^{33}$
\and         K. Matsumoto\orcit{0000-0002-5277-568X}$^{41}$
\and      W. Max-Moerbeck$^{42}$
\and             B. Mihov$^{13}$
\and             M. Minev\orcit{0000-0002-5702-5095}$^{13}$
\and      M. G. Mingaliev\orcit{0000-0001-8585-1186}$^{43,44,10}$
\and         A. Modaressi$^{16}$
\and       D. A. Morozova\orcit{0000-0002-9407-7804}$^{26}$
\and           F. Mortari$^{23}$
\and     T. V. Mufakharov\orcit{0000-0001-9984-127X}$^{43,44,36}$
\and          I. Myserlis\orcit{0000-0003-3025-9497}$^{45,37}$
\and   M. G. Nikolashvili$^{ 2}$
\and        T. J. Pearson$^{32}$
\and         A. V. Popkov\orcit{0000-0002-0739-700X}$^{46,35,36}$
\and        I. A. Rahimov\orcit{0000-0002-9185-6239}$^{10,11}$
\and    A. C. S. Readhead$^{32}$
\and          D. Reinhart$^{21}$
\and            R. Reeves$^{47}$
\and           S. Righini\orcit{0000-0001-7332-5138}$^{25}$
\and        F. D. Romanov\orcit{0000-0002-5268-7735}$^{48,49,50}$
\and      S. S. Savchenko\orcit{0000-0003-4147-3851}$^{26,51}$
\and            E. Semkov\orcit{0000-0002-1839-3936}$^{13}$
\and      E. V. Shishkina\orcit{0009-0002-2440-2947}$^{26}$
\and          L. A. Sigua$^{ 2}$
\and  L. Slavcheva-Mihova$^{13}$
\and     Yu. V. Sotnikova\orcit{0000-0001-9172-7237}$^{43,44,36}$
\and          R. Steineke$^{21}$
\and        M. Stojanovic\orcit{0000-0002-4105-7113}$^{19}$
\and        A. Strigachev$^{13}$
\and             A. Takey\orcit{0000-0003-1423-5516}$^{30}$
\and          E. Traianou\orcit{0000-0002-1209-6500}$^{ 9}$
\and    Yu. V. Troitskaya\orcit{0000-0002-9907-9876}$^{26}$
\and      I. S. Troitskiy\orcit{0000-0002-4218-0148}$^{26}$
\and           A. L. Tsai\orcit{0000-0002-3211-4219}$^{18,52}$
\and          A. Valcheva\orcit{0000-0002-2827-4105}$^{15}$
\and       A. A. Vasilyev\orcit{0000-0002-8293-0214}$^{26}$
\and             G. Verna$^{40}$
\and             O. Vince\orcit{0009-0008-5761-3701}$^{19}$
\and          K. Vrontaki\orcit{0009-0002-7669-7425}$^{24}$
\and         Z. R. Weaver\orcit{0000-0001-6314-0690}$^{33}$
\and              J. Webb\orcit{0000-0001-6078-2022}$^{53}$
\and      Q. X. Yuldoshev\orcit{0000-0002-6554-3618}$^{ 3}$
\and         E. Zaharieva\orcit{0000-0001-7663-4489}$^{15}$
\and        A. V. Zhovtan$^{14}$
 }

\institute{
INAF, Osservatorio Astrofisico di Torino, via Osservatorio 20, I-10025 Pino Torinese, Italy \email{claudia.raiteri@inaf.it}
\and                                                                                                                                          Abastumani Observatory, Mt. Kanobili, 0301 Abastumani, Georgia
\and                                                                                                                      Ulugh Beg Astronomical Institute, Astronomy Street 33, Tashkent 100052, Uzbekistan
\and                                                                                                   Instituto de Astronom\'ia, Universidad Nacional Aut\'onoma de M\'exico, AP 70-264, CDMX 04510, Mexico
\and                                                                                                               INAF, Osservatorio Astronomico di Brera, via Emilio Bianchi 46, 23807, Merate (LC), Italy
\and                                                                                                                                                            EPT Observatories, Tijarafe, La Palma, Spain
\and                                                                                                                                                  INAF, TNG Fundaci\'on Galileo Galilei, La Palma, Spain
\and                                                                                                                  Instituto de Astrof\'isica de Canarias, Via Lactea, E-38200 La Laguna, Tenerife, Spain
\and                                                                                                    Instituto de Astrof\'isica de Andaluc\'ia-CSIC, Glorieta de la Astronom\'ia, E-18008, Granada, Spain
\and                                                                                                            Institute of Applied Astronomy of RAS, Kutuzova Embankment 10, St.~Petersburg 191187, Russia
\and                                                                                                                                                     Radioastronomical observatory “Svetloe”, Russia
\and                                                                                                            Brigham Young University Department of Physics and Astronomy, N284 ESC, Provo, UT 84602, USA
\and                                                         Institute of Astronomy and National Astronomical Observatory, Bulgarian Academy of Sciences, 72 Tsarigradsko shosse Blvd., 1784 Sofia, Bulgaria
\and                                                                                                                                      Crimean Astrophysical Observatory RAS, P/O Nauchny, 298409, Russia
\and                                                                   Department of Astronomy, Faculty of Physics, Sofia University "St. Kliment Ohridski", 5 James Bourchier Blvd, BG-1164 Sofia, Bulgaria
\and                                                                                                                                              Connecticut College, 270 Mohegan Ave., New London, CT, USA
\and                                                                                                      Institute of Astrophysics, Foundation for Research and Technology - Hellas, 7110 Heraklion, Greece
\and                                                                                                                           National Central University, 300 Zhongda Road, Zhongli 32001, Taoyuan, Taiwan
\and                                                                                                                                             Astronomical Observatory, Volgina 7, 11060 Belgrade, Serbia
\and                                                                                                                                          National University of Uzbekistan, Tashkent 100174, Uzbekistan
\and                                                             Hans-Haffner-Sternwarte (Hettstadt), Naturwissenschaftliches Labor f\"ur Sch\"uler, Friedrich-Koenig-Gymnasium, D-97082 W\"urzburg, Germany
\and                                                                                                                        Astroteilchenphysik, TU Dortmund, Otto-Hahn-Str. 4A, D-44227 Dort­mund, Germany
\and                                                                                                                                    Hypatia Observatory, 19 Via Sacco e Vanzetti, Viserba, Rimini, Italy
\and                                                           Section of Astrophysics, Astronomy and Mechanics, Dept. of Physics, National and Kapodistrian Univ. of Athens, 15784 Zografos, Athens, Greece
\and                                                                                                                             INAF – Istituto di Radioastronomia, Via Gobetti 101, 40129 Bologna, Italy
\and                                                                                                            Saint Petersburg State University, 7/9 Universitetskaya nab., St.~Petersburg, 199034, Russia
\and                                                                                                    Aryabhatta Research Institute of Observational Sciences (ARIES), Manora Peak, Nainital 263001, India
\and                                                                                                                      Xinjiang Astronomical Observatory, CAS, 150 Science-1 Street, Urumqi 830011, China
\and                                                                                              Center for Astrophysics \text{\textbar} Harvard \& Smithsonian, 60 Garden Street, Cambridge, MA 02138, USA
\and                                                                                                             National Research Institute of Astronomy and Geophysics (NRIAG), 11421 Helwan, Cairo, Egypt
\and                                                                                           Instituto de Astronom\'ia, Universidad Nacional Aut\'onoma de M\'exico, AP 106, Ensenada 22800, B. C., Mexico
\and                                                                                                             Owens Valley Radio Observatory, California Institute of Technology, Pasadena, CA 91125, USA
\and                                                                                                                                    IAR, Boston University, 725 Commonwealth Ave., Boston, MA 02215, USA
\and                                                                                                                                  Department of Physics, University of Crete, GR-70013 Heraklion, Greece
\and                                                                                                            Astro Space Center of Lebedev Physical Institute, Profsoyuznaya 84/32, 117997 Moscow, Russia
\and  Institute for Nuclear Research, Russian Academy of Sciences, 60th October Anniversary Prospect 7a, Moscow 117312, Russia
\and                                                                                                                      Max-Planck-Institut f\"ur Radioastronomie, Auf dem H\"ugel 69, 53121 Bonn, Germany
\and                                                                                                                        Engelhardt Astronomical Observatory, Kazan Federal University, Tatarstan, Russia
\and                                                                                               Facultad de Ciencias, Universidad Aut\'onoma de Baja California, El Sauzal, Ensenada 22800,  B.C., Mexico
\and                                                                                                                         Astronomical Observatory, University of Siena, Via Roma 56, 53100, Siena, Italy
\and                                                                                                                                  Astronomical Institute, Osaka Kyoiku University, Osaka 582-8582, Japan
\and                                                                                            Departamento de Astronom\'ia, Universidad de Chile, Camino El Observatorio 1515, Las Condes, Santiago, Chile
\and                                                                                                                                 Special Astrophysical Observatory of RAS, Nizhny Arkhyz, 369167, Russia
\and                                                                                                                                     Kazan Federal University, 18 Kremlyovskaya St, Kazan 420008, Russia
\and                                                                                                    Institut de Radioastronomie Milim\'etrique, Avenida Divina Pastora 7, Local 20, 18012 Granada, Spain
\and                                                                                                             Moscow Institute of Physics and Technology, Institutsky per. 9, Dolgoprudny, 141700, Russia
\and                                                                                                                   CePIA, Departamento de Astronom\'ia, Universidad de Concepci\'on, Concepci\'on, Chile
\and                                                                                                                                                 American Association of Variable Star Observers (AAVSO)
\and                                                                                                     Burke-Gaffney Observatory. Saint Mary’s University, 923 Robie Street, Halifax, NS B3H 3C3, Canada
\and                                                                                                                            Abbey Ridge Observatory, 45 Abbey Road, Stillwater Lake, Nova Scotia, Canada
\and                                                                                                                                                      Pulkovo Observatory, St.~Petersburg, 196140, Russia
\and                                                                                                                                National Sun Yat-sun University, 70 Lienhai Rd., Kaohsiung 80424, Taiwan
\and                                                                                                                       Florida International University, the SARA observatories, FIU, Miami Florida, USA
 }
 
   \date{}

% \abstract{}{}{}{}{} 
% 5 {} token are mandatory
 
  \abstract
   {Blazars are beamed active galactic nuclei (AGNs) known for their strong multi-wavelength variability on timescales ranging from years down to minutes. Many different models have been proposed to explain this variability.}
   {We aim to investigate the suitability of the twisting jet model presented in previous works to explain the multi-wavelength behaviour of BL Lacertae, the prototype of one of the blazar classes. According to this model, the jet is inhomogeneous, curved, and twisting, and the long-term variability is due to changes in the Doppler factor due to variations in the orientation of the jet-emitting regions.}
   {We analysed optical data of the source obtained during monitoring campaigns organised by the Whole Earth Blazar Telescope (WEBT) in 2019--2022, together with radio data from the WEBT and other teams, and $\gamma$-ray data from the {\it Fermi} satellite. In this period, BL Lacertae underwent an extraordinary activity phase, reaching its historical optical and $\gamma$-ray brightness maxima.}
   {The application of the twisting jet model to the source light curves allows us to infer the wiggling motion of the optical, radio, and $\gamma$-ray jet-emitting regions. The optical-radio correlation shows that the changes in the radio viewing angle follow those in the optical viewing angle by about 120 days, and it suggests that the jet is composed of plasma filaments, which is in agreement with some radio high-resolution observations of other sources. The $\gamma$-ray emitting region is found to be co-spatial with the optical one, and the analysis of the $\gamma$-optical correlation is consistent with both the geometric interpretation and a synchrotron self-Compton (SSC) origin of the high-energy photons.}
   {We propose a geometric scenario where the jet is made up of a pair of emitting plasma filaments in a sort of double-helix curved rotating structure, whose wiggling motion produces changes in the Doppler beaming and can thus explain the observed multi-wavelength long-term variability.}


   \titlerunning{A wiggling filamentary jet at the origin of blazar behaviour}
   \authorrunning{C. M. Raiteri et al.}

   \maketitle
%
%-------------------------------------------------------------------

\section{Introduction}
Blazars are peculiar active galactic nuclei (AGNs) that are characterised by a relativistic jet pointing at a small angle with respect to the line of sight. This results in Doppler beaming of the emitted radiation, with consequent flux enhancement and shortening of the variability timescales; blazars thus appear as strongly variable objects at all wavelengths on a variety of timescales.

Many scenarios have been proposed to explain the unpredictable blazar variability. They can be categorised into two main classes. The first class involves intrinsic energetic processes occurring inside the jet, such as shock waves propagating in the jet \citep[e.g.][]{marscher1985}, magnetic reconnection \citep[e.g.][]{sironi2015,petropoulou2018,bodo2021}, and turbulence \citep[e.g.][]{marscher2014}.
The second class involves changes in beaming due to geometric mechanisms, such as orbital motion in a binary black hole system \citep[e.g.][]{lehto1996,villata1998} or jet precession \citep[e.g.][]{britzen2018}. These two mechanisms would cause periodic behaviour, but robust evidence of persistent periodicity has only been found in a few cases.

An alternative geometric model was proposed by \citet{raiteri2017_nature} to explain the long-term blazar variability. According to this model, the jet is inhomogeneous, curved, and twisting. 'Inhomogeneous' means that emission at different frequencies comes from different regions of the jet. 
In particular, synchrotron radiation of increasing frequency comes from increasingly further in jet regions.
`Curved' implies that these regions have different orientations relatively to one another and therefore different beaming, because beaming depends on the viewing angle, that is the angle between the plasma velocity and the line of sight.
`Twisting' means that the orientation of the various emitting zones changes over time, as does the amount of beaming affecting the radiation they produce. 
There are many observations \citep[e.g.][]{owen1980,giroletti2004,perucho2012,fromm2013,casadio2015,britzen2017,britzen2018,lister2021,issaoun2022,zhao2022,pushkarev2023} and theoretical results \citep[e.g.][]{begelman1980,nakamura2001,hardee2003,moll2008,mignone2010,liska2018} that suggest changes in the jet orientation in both space and time, giving support to the twisting jet model. This model was successfully applied to the behaviour of several blazars observed by the Whole Earth Blazar Telescope\footnote{\url{https://www.oato.inaf.it/blazars/webt/}} (WEBT) collaboration during its monitoring campaigns \citep[e.g.][]{villata2002,villata2007,raiteri2013,raiteri2017_nature,raiteri2021b}. 
However, the origin of the irregular twisting motion remains unclear.
%, and it has been ascribed to either the classical geometric variability mechanisms mentioned above, or to plasma magnetohidrodinamical instabilities developing in the jet (ref), or to a combination of both.

Further clues in this regard have come from the detection of many characteristic variability timescales, from sub-days to several days, in the light curves of blazars observed with exceptionally high temporal resolution by the Transiting Exoplanet Survey Satellite (TESS). This was interpreted as the contribution of several wrapped filaments forming a twisting jet \citep{raiteri2021a,raiteri2021b}.
The twisted filamentary structure of extragalactic jets has been known for many years from radio, optical, and ultraviolet images of the radio galaxy M~87 \citep[e.g.][and references therein]{owen1989,boksenberg1992,nikonov2023}. 
Radio interferometric observations have revealed the presence of two helical threads inside the jet of the blazar 3C~273 \citep{lobanov2001}.
A twisted filamentary jet structure was recently observed in 3C~279 and proposed as an explanation for the blazar radio variability \citep{fuentes2023}.
The extremely well-resolved images acquired with RadioAstron and supporting ground antennas allowed the authors to recognise that the emission comes from two - maybe three - interlaced rotating filaments, possibly produced by plasma instabilities. They suggested that the brightest zones that appear to move along the filaments are due to enhanced Doppler boosting because of a better alignment with the line of sight, as already suggested by, for instance, \citet{bach2006} and \citet{raiteri2021b}. This points towards a geometric interpretation of blazar variability in line with the twisting-jet model.
%The authors analysed radio images at 22 GHz acquired by the RadioAstron mission and suggested that plasma instabilities in a rotating jet with a helical magnetic field give rise to thin emitting filaments, with variability time scales that can be much shorter than those predicted by causality arguments applied to the whole jet, thus explaining the vary fast variability with minute time scales detected at various frequencies (ref).
Confirming this picture through photometric data requires exceptionally well-sampled light curves, which is only possible by means of a wide network of telescopes such as the WEBT. 

The object BL Lacertae is a bright and very variable source and is an ideal candidate to study also because its sub-parsec jet shows a wiggling helical structure that changes in time \citep{cohen2015,kim2023}.
This source recently underwent an extraordinary activity phase, %reaching its maximum optical brightness level, 
which was continuously monitored by the WEBT Collaboration \citep{jorstad2022,raiteri2023a}. 
%The detection of quasi-periodic oscillations in the optical and $\gamma$-ray flux and optical polarization during the first part of the 2020 outburst led \citet{jorstad2022} to suggest the development of kink instabilities triggered by a perturbation travelling in the jet.
%\citet{raiteri2023} analysed the optical behaviour in the 2021--2022 observing season and found several clues in favour of the twisting jet interpretation of the long-term flux and polarization variability. 
In this work, we tested the hypothesis of a filamentary twisting jet to explain the multi-wavelength behaviour of BL Lacertae by analysing optical, radio, and $\gamma$-ray data over a wide period, ranging from 2019 January 1 to 2022 February 28.
The optical and radio data are presented in Sects.~\ref{sec:optical} and \ref{sec:radio}, respectively. The application of the twisting-jet model to the optical and radio emission is discussed in Sect.~\ref{sec:or}. In Sect.~\ref{sec:dcf_or}, we analyse the optical-radio correlation and propose an interpretation in terms of emitting jet filaments. 
A description of the proposed wiggling filamentary jet model is given in Sect.~\ref{sec:model}, together with a graphical representation. 
The $\gamma$-ray data are presented in Sect.~\ref{sec:gamma}, and their correlation with the optical data is analysed in Sect.~\ref{sec:dcf_go}. The twisting-jet model is applied to the $\gamma$-ray fluxes in Sect.~\ref{sec:go}. Tests of the model are discussed in Sect.~\ref{sec:tests}. Sect.~\ref{sec:end} contains a summary and a final discussion.

%--------------------------------------------------------------------
\section{Optical observations}
\label{sec:optical}

   \begin{figure*}
   \sidecaption
   \includegraphics[width=12cm]{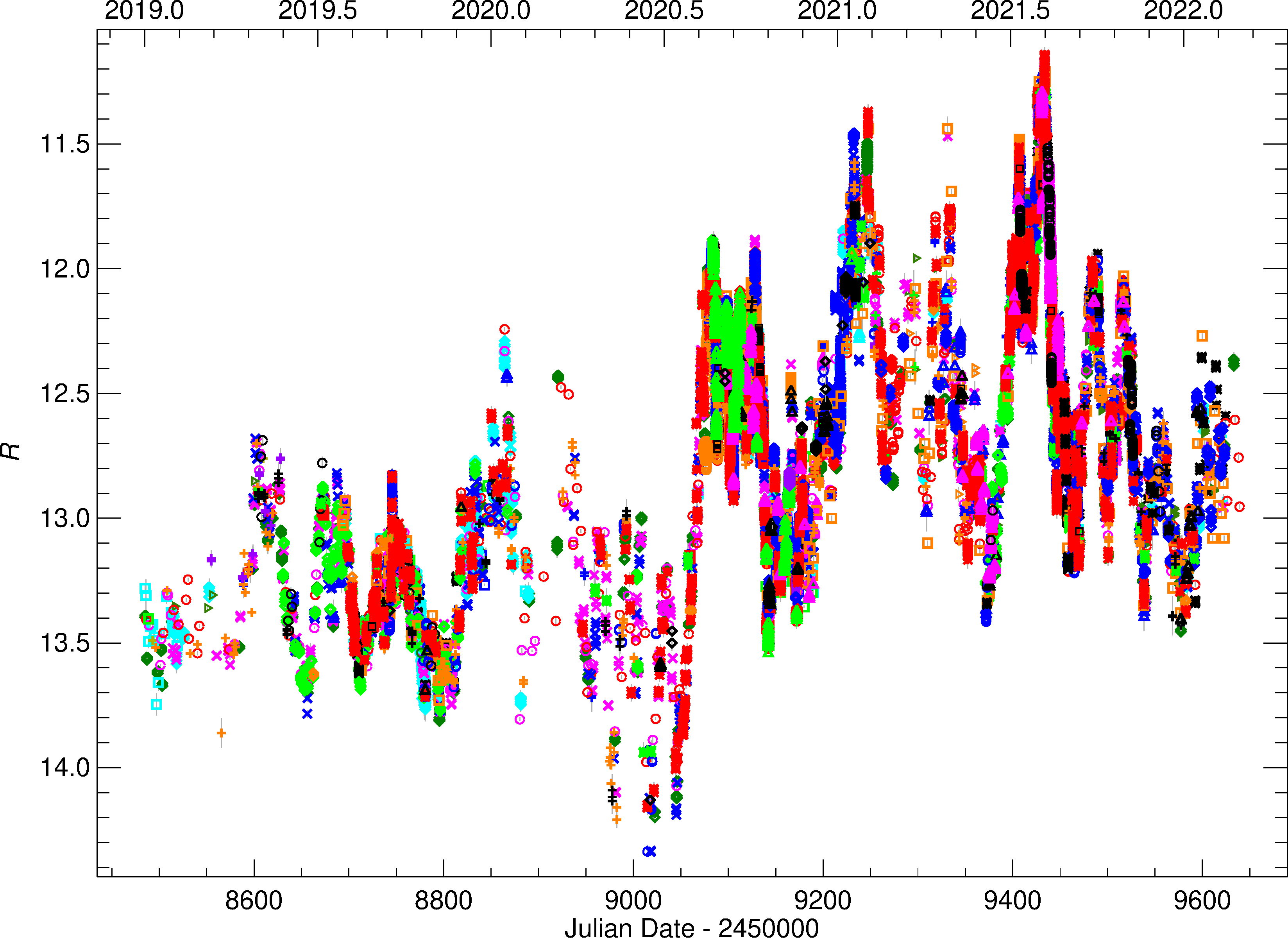}
   \caption{Optical light curve (observed magnitudes) of BL~Lacertae in the $R$ band  obtained by the WEBT in 2019--2022. Different symbols and colours distinguish between the various datasets as indicated in Table~\ref{tab:optical}.}
   \label{fig:webt}%
   \end{figure*}

The observations analysed in this paper cover the period from 2019 January 1 (JD=2458484.5) to 2022 February 28 (JD=2459639.5). They were performed in the framework of subsequent WEBT campaigns on BL Lacertae.
 
In the optical band, we assembled data in the $R$ band provided by 57 datasets from 48 observatories in 17 countries (see Table~\ref{tab:optical}). To obtain as homogeneous as possible photometry, all observers adopted the same prescriptions to derive the source standard magnitude. This was obtained with respect to the photometric sequence published by \citet{fiorucci1996} and using a common aperture radius of 8 arcsec, with background extracted in a surrounding annulus of 10 and 16 arcsec radii.
We processed the data in order to obtain a reliable $R$-band light curve, which is shown in Fig.~\ref{fig:webt}, containing a total of 35074 data points.
%vedi conta.ndata
Part of them were published in \citet{weaver2020}, \citet{jorstad2022}, and \citet{raiteri2023a}.

   During the considered period, the source showed continuous flaring activity, with a maximum amplitude of about 3.2 mag.
   We can distinguish a lower brightness state before $\rm JD \sim 2459050$ (2020 July 19), where the mean magnitude is $R \approx 13.36$, and a subsequent brighter state with mean magnitude $R \approx 12.46$. In this brighter state the source reached its historical maximum $R=11.14 \pm 0.03$ on 2021 August 7 ($\rm JD = 2459433.7$).

We converted magnitudes into flux densities, correcting for both Galactic extinction (0.713 mag from the NASA/IPAC Extragalactic Database - NED\footnote{\url{https://ned.ipac.caltech.edu/}}) and host-galaxy contribution (about 2.5 mJy for our photometric prescriptions, see \citealt{raiteri2009}).
%0.60*4.229
Throughout the paper, we always use these corrected values.

\section{Radio observations}
\label{sec:radio}

\begin{figure}
   \resizebox{\hsize}{!}{\includegraphics{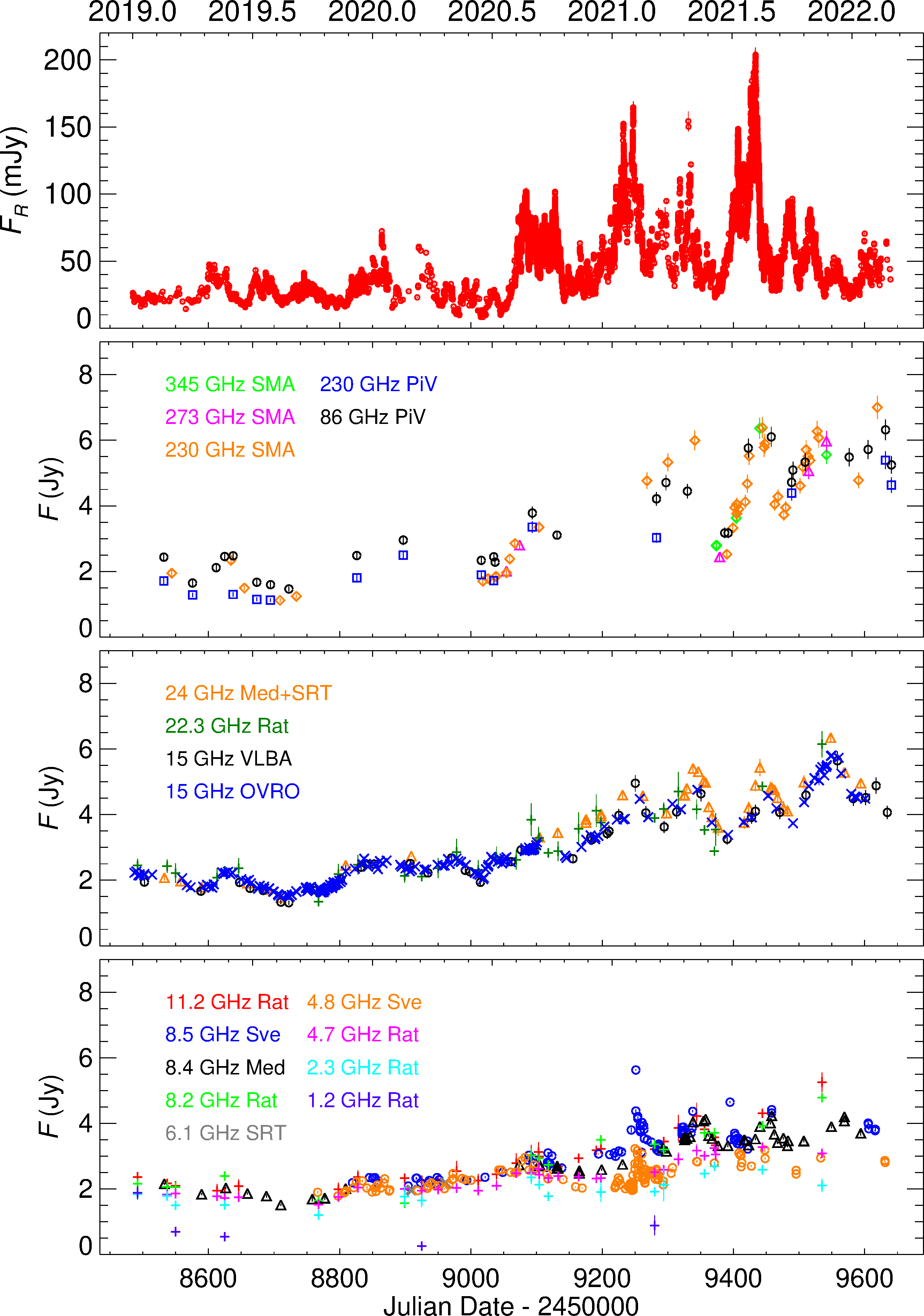}}
   \caption{Optical flux densities of BL Lacertae in $R$ band (top) and radio flux densities at different frequencies. Different symbols and colours distinguish between the various radio datasets as indicated in Table~\ref{tab:radio}.}
   \label{fig:radop}
\end{figure}

In the radio band, we collected 18 datasets from eight facilities; these are listed in Table~\ref{tab:radio}. 
Radio data were acquired in different bands, spanning from 1.2 GHz to 345 GHz.
The Very Long Baseline Array (VLBA) data at 15 GHz are from the MOJAVE program\footnote{\url{https://www.cv.nrao.edu/MOJAVE/}} \citep{lister2018} and represent total cleaned Stokes I flux densities in the VLBA images. 
Data from the Owens Valley Radio Observatory (OVRO) were acquired and reduced as explained in \citet{richards2011}.
The radio light curves are shown in Fig.~\ref{fig:radop} and compared to the optical flux densities in the $R$ band.

The best-sampled radio light curve is the one at 15 GHz, obtained by combining data from the OVRO and MOJAVE programmes. There is no significant offset in the fluxes from these two datasets, which would instead be expected in the case of diffuse emission.
%CHECK scaling between the two datasets
The light curve shows a variability amplitude of a factor of $\sim 4.4$, from $F=1.31 \, \rm Jy$ to $F=5.80 \, \rm Jy$, and a mean value of $2.70 \, \rm Jy$.
Radio flux densities at 22--24 GHz confirm the trend traced by the data at 15 GHz, with slightly larger variability amplitude.
The radio light curves at higher frequencies are much less sampled than the 15 GHz one; they display a behaviour which is consistent with that observed at 15 GHz, with some larger variability amplitude.
Going towards the longest wavelengths, the light curves flatten, and only from 2021.0 onward do they show some enhanced variability.

%Radio observations were provided by:
%the SubMillimeter Array in US (SMA, 345, 273, and 230 GHz), an array of 8 dishes of 6~m each,
%the 30~m IRAM telescope in Pico Veleta, Spain (at 230 and 86 GHz),  
%the 14~m radio telescope in Mets\"ahovi, Finland (37 GHz), 
%the 32~m and 64~m radio telescopes in Medicina and Sardinia, Italy (24, 8.4 and 6.1 GHz), 
%the 40~m telescope of the Owens Valley Radio Observatory in US (OVRO, 15 GHz),
%the Very Long Baseline Array (VLBA) calibrators database (15 GHz), 
%the RATAN-600 radio telescope in Russia (22, 11, 8.2, 4.7, 2.3, and 1.2 GHz), a ring with a diameter of 576~m, and  the Svetloe 32~m radio telescope (8.5 and 4.8 GHz).

\section{Twisting-jet model applied to the optical and radio data}
\label{sec:or}
According to the twisting jet model of \citet{raiteri2017_nature}, the long-term trend of the flux at a given wavelength is the result of variations in the Doppler factor due to changes in the viewing angle of the corresponding emitting region in the jet. A better alignment with respect to the line of sight implies a higher Doppler factor, and hence an enhanced observed flux.

\begin{figure}
   \resizebox{\hsize}{!}{\includegraphics{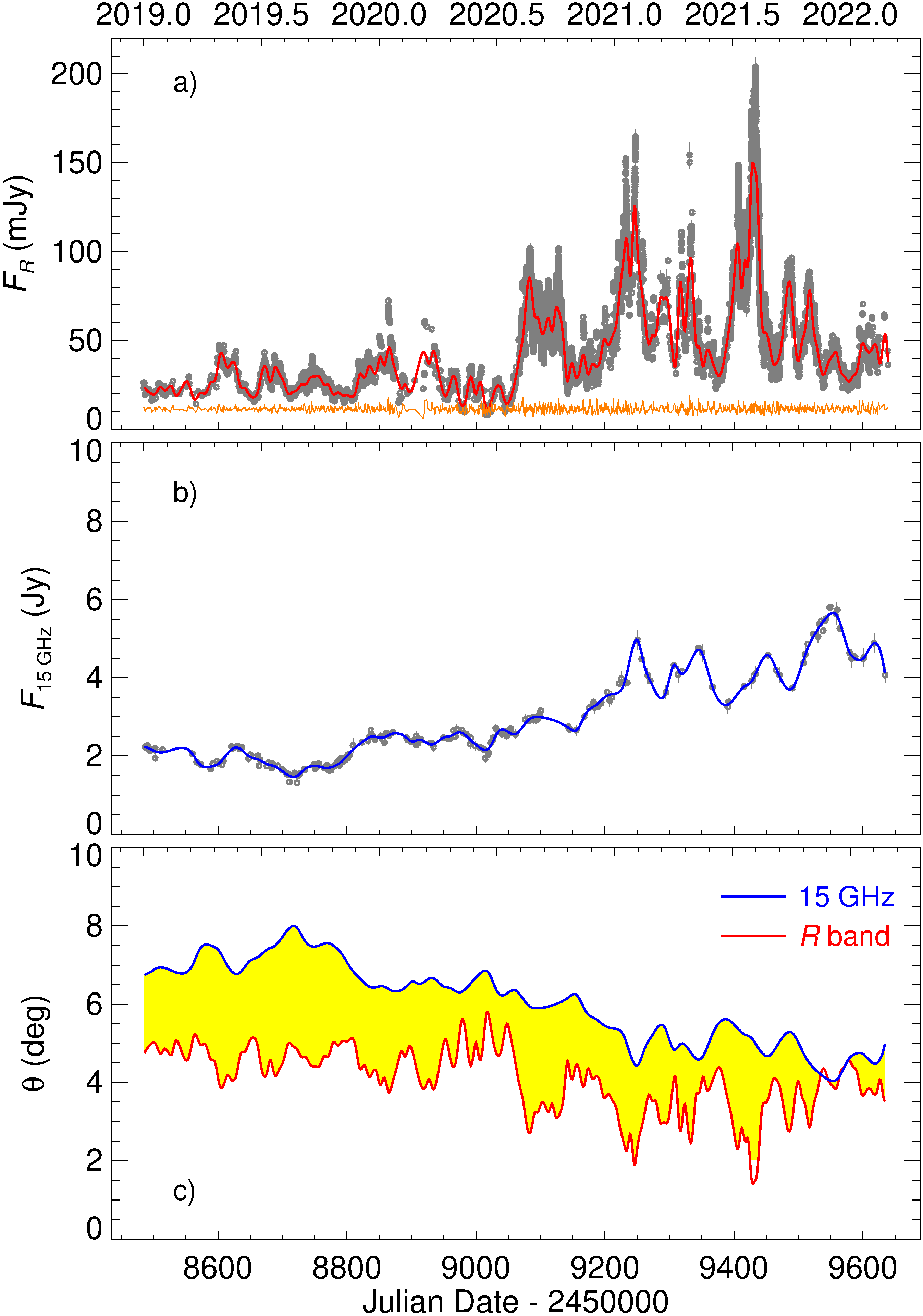}}
   \caption{Flux densities in $R$ band (a) and at 
   %37 GHz % con Metsahovi
   15 GHz (b) and behaviour in time of viewing angle (c) of optical-emitting region (red) and radio-emitting region (blue). The red and blue lines in panels (a) and (b) show cubic spline interpolations on the binned optical and radio data, respectively. They represent the long-term trends of the flux. Deboosted optical flux densities are displayed in orange; they represent fluxes corrected for the effect of variable Doppler beaming. 
In the bottom panels, the yellow areas highlight the periods where the optical-emitting region is better aligned with the line of sight than the radio zone, and hence the optical radiation is more beamed than the radio emission.}
   \label{fig:radop_teta}%
   \end{figure}

The Doppler factor is defined as
\begin{equation}
\delta=[\Gamma \, (1-\beta \, \cos \theta)]^{-1},
\label{eq:delta}
\end{equation}
where $\theta$ is the viewing angle, $\beta$ is the bulk velocity of the plasma in the jet in units of the speed of light, and $\Gamma=(1-\beta^2)^{-1/2}$ is the bulk Lorentz factor. Under the assumption that the long-term trend of the flux is due to variations of the Doppler factor, we can derive the optical and radio long-term trends by interpolating cubic splines through the flux densities in the $R$ band and at 15 GHz, following the method discussed in \citet{raiteri2017_nature}.
The resulting long-term trends are shown in Fig.~\ref{fig:radop_teta}.
Then, we can obtain the behaviour of the Doppler factor in time by reversing the relation
\begin{equation}
F_\nu (t) \propto \delta^{n+\alpha} (t) \, ,
\label{eq:flux}
\end{equation}
where $n=2$ for a continuous jet \citep{urry1995} and $\alpha$ is the spectral index of the power law $F_\nu (\nu) \propto \nu^{- \alpha}$.
We fixed $\alpha=2$ for the optical spectrum \citep{raiteri2023a} and $\alpha=0$ for the radio one, as inferred from the data (see Sect.~\ref{sec:radio}).
The Lorentz factor was set to $\Gamma=10$, a typical value for BL Lac objects \citep[e.g.][]{hovatta2009}.
To normalise Eq.~\ref{eq:flux}, we need to find, for instance, the minimum $\delta$ corresponding to the minimum long-term flux density (and thus to the maximum viewing angle), $\delta_{\rm min}=[\Gamma (1-\beta \cos\theta_{\rm max})]^{-1}$.  
This requires fixing $\theta_{\rm max}$.
In order to put the data analysed in this paper in a historical context, we explored the optical and radio behaviour of BL Lacertae in the past, exploiting the data collected during previous WEBT campaigns \citep{villata2002,villata2004a,villata2009,raiteri2009,raiteri2010,raiteri2013} and stored in the WEBT archive. These data cover the 1993--2012 period and are shown in Fig.~\ref{fig:radop_teta_past} in the appendix. 
%The optical flux densities are much lower than those in the recent period displayed in Fig.~\ref{fig:radop_teta}. 

At 15 GHz, the minimum flux is found in the recent light curve (see Fig.~\ref{fig:radop_teta}), and we attributed it to $\theta_{\rm max}=8^\circ$. 
In the optical, the source was in general much fainter and reached a lower flux minimum in 1993--2012 than in the 2019--2022 period. The maximum optical viewing angle was set in a way to obtain the same mean orientation of the optical- and radio-emitting regions in the nearly 20 years of the historical record; so, $\theta_{\rm max} \approx 8.4^\circ$ for the optical zone.
%$\theta_{\rm max}=8.36^\circ$
Once we have the Doppler factor as a function of time, we can derive the behaviour of the viewing angle in time by reversing Eq.~\ref{eq:delta}. The result is shown in Fig.~\ref{fig:radop_teta} (see Fig.~\ref{fig:radop_teta_past} in the appendix for the historical light curves).
The variation in time of the viewing angle allows us to obtain a picture of the twisting motion of the jet at the location of the zones producing the optical and radio radiation. The optical region appears to wobble faster, which is expected if it is smaller than the radio-emitting zone, as suggested by the general faster variability of the optical flux. The smoother behaviour of the radio flux and viewing angle is due to the fact that we observe average values over a much larger emitting zone. 

In the considered period, the optical viewing angle is almost always smaller than the radio one, with consequent stronger Doppler beaming of the optical radiation with respect to the radio emission.
However, the viewing angle of the radio-emitting region progressively comes closer to that of the optical zone, and in the last part of the period, corresponding to the big outbursts, the two emitting regions achieve the best alignment with the line of sight.

It is worth mentioning that with the choice of parameters described above, the Doppler factor that we inferred for the 15 GHz radiation ranges between about 7 and 15, and the corresponding viewing angle goes from about $3^\circ$ to $8^\circ$.
These values are comparable to those obtained with different methods using radio data from Mets\"ahovi and MOJAVE
 (\citealt{savolainen2010}: $\delta = 7.2$, $\Gamma=5.4$, $\theta=7.5$), from OVRO and MOJAVE  (\citealt{liodakis2018}: $\delta \sim 12$, $\Gamma \sim 10$, $\theta \sim 4.6$), or from MOJAVE alone (\citealt{homan2021}: $\delta=18.3$, $\Gamma=11.9$, $\theta=2.6$).
 On the optical side, we obtained $\delta$ values in the 6.4--19.2 range and viewing angles from 1.2$^\circ$ to 8.4$^\circ$.
 The minimum optical viewing angle corresponds to the peak of the 2021 outburst (see Fig.~\ref{fig:radop_teta}), and its value is in agreement with the results in \citet{raiteri2023a}. 
 %Indeed, we expect that the jet is closely aligned with the line of sight at a maximum brightness level.

We note that knowledge of the Doppler factor behaviour in time allowed us to correct the observed fluxes for the variable beaming effect and to derive the `deboosted' fluxes \citep[see][]{raiteri2017_nature}. 
These represent the jet emission in the case of a constant Doppler factor; so, they differ only by a scale factor from the intrinsic flux densities. Their short-term variability is likely due to energetic processes in the jet.
The deboosted optical fluxes are shown in the top panel of Fig.~\ref{fig:radop_teta} and are further discussed in Sect.~\ref{sec:tests}.

\section{Correlation between the optical and radio emission}
\label{sec:dcf_or}
We investigated the possible correlation between the orientation changes in the optical and radio jet zones by means of the discrete correlation function \citep[DCF;][]{edelson1988,hufnagel1992}. This quantifies the strength of the correlation $r$ between two unevenly sampled time series as a function of the time lag $\tau$.
We ran the DCF on cubic-spline interpolations through the light curves in the $R$ band and at 15 GHz. These splines represent the long-term trends from which we derived the behaviour of the viewing angles (see Fig.~\ref{fig:radop_teta}) and can therefore be considered as proxies for $\theta$.

   \begin{figure}
   \resizebox{\hsize}{!}{\includegraphics{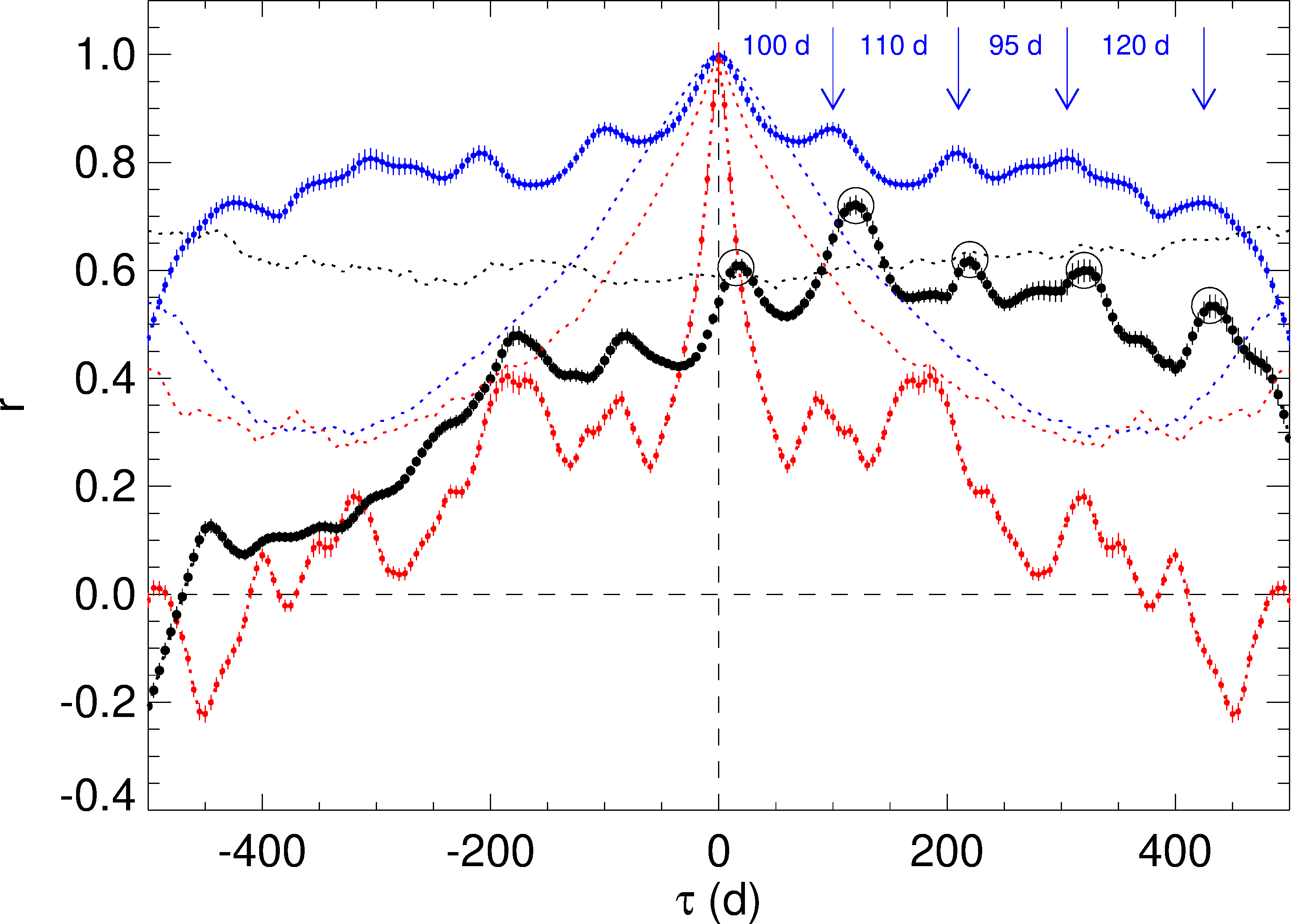}}
      \caption{DCF between optical and radio cubic spline interpolations to corresponding light curves, used as proxies for viewing angles (black dots). Large empty circles highlight the highest DCF peaks. The blue and red dots represent the ACFs of the radio and optical splines, respectively. The vertical arrows mark the radio ACF peaks, which are separated by time lags $\tau$ of 95--120 d.
      Dotted lines represent 90\% confidence levels of the ACFs and DCF of the same colour, which were obtained by cross-correlating splines on 1000 optical and 1000 radio simulated light curves, according to the method discussed in \cite{emma2013}.}
         \label{fig:dcf_or}
   \end{figure}
The results are shown in Fig.~\ref{fig:dcf_or}, which also displays the 90\% confidence levels derived by running the DCF on the cubic spline interpolations through 1000 artificial optical and 1000 artificial radio light curves obtained with the method developed in \citet{emma2013}.
The auto-correlation function (ACF) resulting from the cross-correlating of the radio spline with itself reveals a series of bumps well above the 90\% confidence level. They are separated by alternating time intervals of 95--100 d and 110-120 d and may be the signature of two plasma filaments contributing to the radio emission, which are alternately crossing the line of sight and producing the flares visible in the radio light curve.
In contrast, the optical ACF only displays peaks below the 90\% confidence level.
The DCF between the optical and radio splines shows that the changes in the viewing angle of the radio-emitting region follow those of the optical-emitting zone.  The most significant peak indicates a time delay of 120 d. Other peaks with decreasing significance are present at time lags of 15, 220, 320, and 430 d. 
%Since the time difference between the optical outbursts is of the order of 100 days, a scenario compatible with the results of the ACF and DCF is that every optical outburst is followed by a radio outburst 120 days after. In other words, every time the optical region better aligns with the line of sight, the radio filaments do the same about four months later. The other DCF signals come from the coupling of different filaments.

A possible scenario that can explain both the features in the DCF and ACF and the behaviour of the light curves is the following. 
In the second half of the optical light curve, we can see four outbursts, which are rather different in shape and brightness; thus, although they are almost equally spaced in time by $\sim 100 \, \rm d$, they do not produce significant signals in the optical ACF.
We may suppose that they are caused by subsequent alignments with the line of sight of the optical zone of the same filaments whose radio-emitting region will align 120 d later, producing the radio outbursts.
In more detail, the first optical outburst in $\sim 2020.7$ is due to the alignment of the first helical filament, which, because of its rotation, causes the first radio outburst $\sim 120 \, \rm d$ later. After an entire turn lasting $\sim 200 \, \rm d$, the same filament produces the third optical outburst, followed by the third radio outburst 120~d later.
The second filament instead gives rise to the second and fourth outbursts in both the optical and the radio light curves, again with a time lag of 120 d of the radio events after the optical ones. Since both the optical and radio outbursts are separated by a time interval of $\sim 100 \, \rm d$, when a radio outburst appears 120~d after its optical counterpart, it has already been preceded by the optical outburst caused by the other filament. This explains the DCF peak at $\tau \sim 15 \, \rm d$.
Similarly, the DCF signals at 220, 320, and 430 d would come from couplings between different filaments or different turns of the same filament.
Finally, we note that a radio delay of $\sim 100 \, \rm d$ after the optical flux variations was already found in \citet{villata2004b} and \citet{villata2009} during previous WEBT campaigns.

%, i.e.~about 100--110 d one after the other, consistent with a possible recurrent alignment of the radio filaments with the optical emitting region. 
%This suggests that 15 d is the time interval necessary for the radio emitting region in the first filament to align with the optical region, while the radio zone in the second filament aligns with the optical region 105 days later. The radio zone in the first filament will align again with the optical region about 100 days later, and so on.

%La periodicita` di circa 100 d si vede solo quando l'angolo di vista e` piccolo (flusso maggiore di 4 Jy), in quanto se l'angolo di vista e` maggiore i due filamenti non sono mai beamati bene e quindi non danno i picchi. Nella curva luce storica infatti questo segnale tra 64 e 128 giorni nel wavelet compare in corrispondenza dei 3 outbursts principali (con determinati settaggi del power). Anche qui, come nell'ottico, si puo` tirare una riga orizzontale a 4 Jy per far vedere che a cio` che spunta sopra i 4 Jy nella curva luce corrisponde un segnale wavelet in quel range 64-128.

%This suggests that the main process producing the variation of the optical zone orientation is also responsible for the change in the alignment of the radio zone.

\section{The wiggling filamentary jet model}
\label{sec:model}

   \begin{figure*}
   \sidecaption
   \includegraphics[width=12cm]{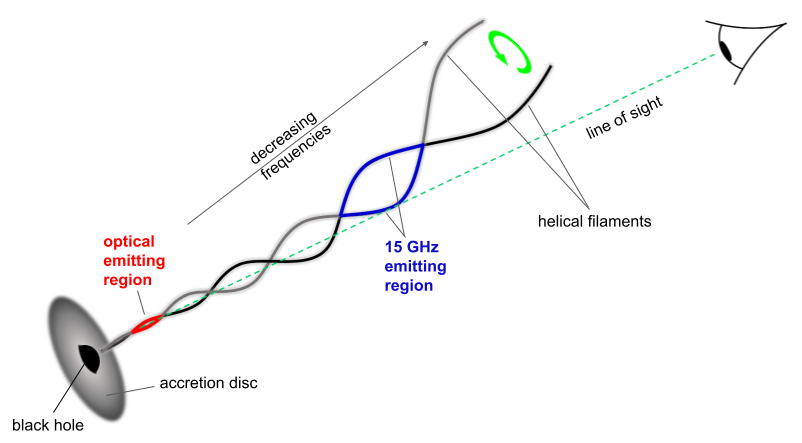}
      \caption{Schematic representation of proposed wiggling filamentary jet model. The jet is inhomogeneous, with radiation of decreasing frequency emitted from increasingly further out jet regions. 
      %The two plasma filaments forming the jet are more tightly wrapped at the location of the optical emitting region, while they are more separated where the radio emission comes from. 
      The radiation coming from the jet regions that have a better alignment with the line of sight undergoes a stronger Doppler beaming. The rotation of the curved double-helix structure produces a time-dependent Doppler boosting in the various jet zones. 
      %It is also responsible for the double delay of the radio flux variations after the optical ones in the period 2019--2022 (see Fig.~\ref{fig:dcf_or}).
              }
         \label{fig:sketch}
   \end{figure*}

According to our analysis, the long-term multi-wavelength behaviour of BL Lacertae can be explained by a model involving a wiggling jet composed of two plasma filaments. 
The filaments can be produced by Kelvin-Helmholtz or current-driven kink instabilities in the plasma (see Sect.~\ref{sec:end}). They give rise to a sort of curved double-helix jet structure, which rotates, producing a variable Doppler beaming of the radiation emitted from the various regions along the jet.
In the inner jet, where the optical emission comes from, the strands are likely tightly wrapped as they are close to the jet apex.
%they are strongly anchored to the accretion disc. 
%Therefore, the optical emission region is rather compact with a single mean viewing angle. 
As we move away downstream, the filaments tend to separate, and at the location where the radio emission is produced, they can be resolved by high-resolution observations such as those presented in \cite{fuentes2023} for 3C~279. 
A schematic representation of the proposed wiggling filamentary jet model is given in Fig.~\ref{fig:sketch}. The model is further discussed in Sect.~\ref{sec:end}.
In Sects.~\ref{sec:gamma}--\ref{sec:go}, we analyse the $\gamma$-ray emission of BL Lacertae in the framework of the above model in order to explore its consistency and consequences.

\section{$\gamma$-ray observations}
\label{sec:gamma}

   \begin{figure}
   \resizebox{\hsize}{!}{\includegraphics{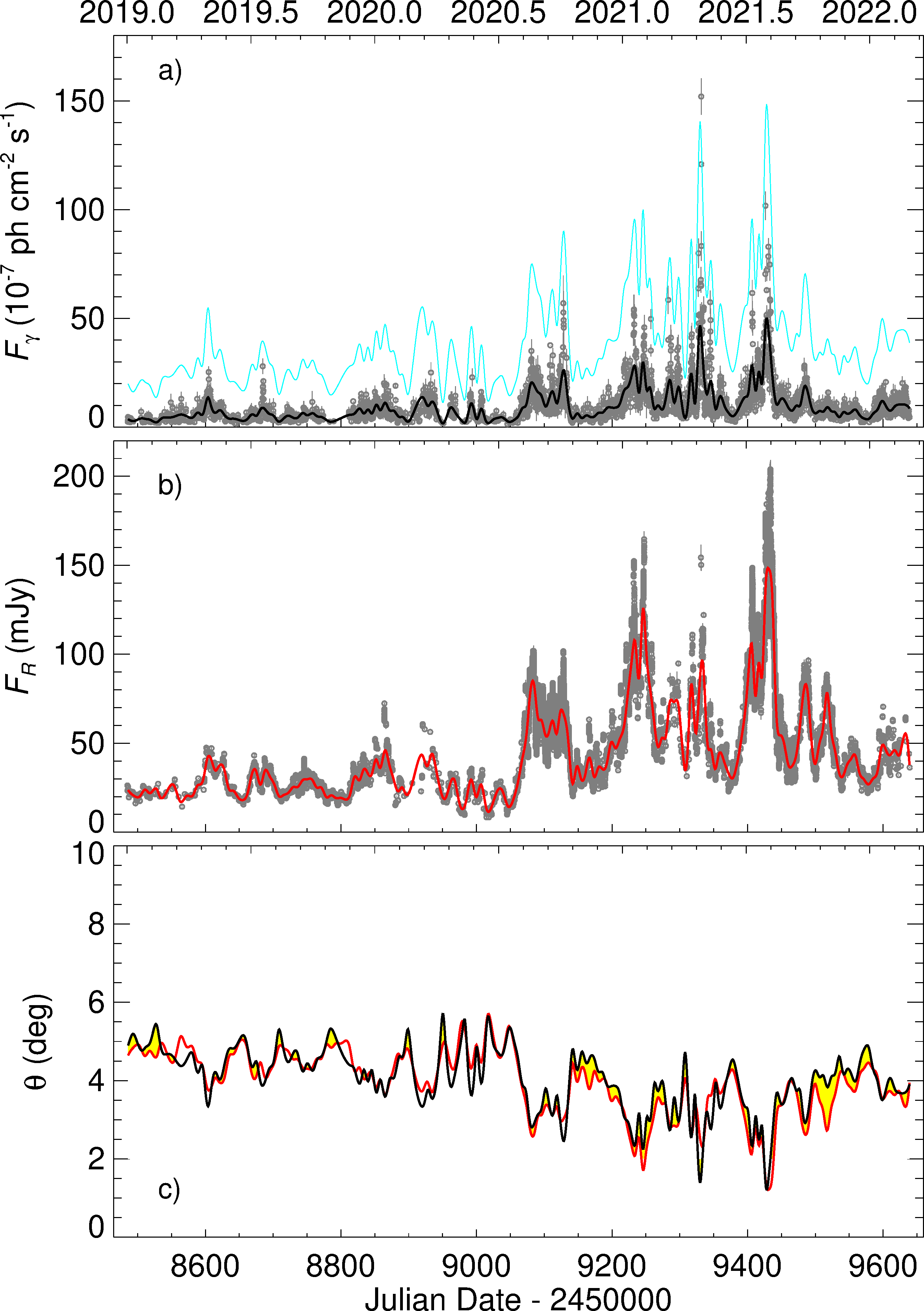}}
   \caption{$\gamma$-ray light curve (a), flux densities in $R$ band (b), and behaviour in time of viewing angle (c) of the $\gamma$-ray emitting region (black) and of the optical emitting region (red).
The black and cyan lines in panel (a) represent the original and rescaled (see Sect.~\ref{sec:go}) cubic spline interpolations on the binned $\gamma$-ray data, respectively; the red line in panel (b) shows the cubic spline interpolation on the binned optical data.
In the bottom panel, the yellow areas highlight the periods where the optical emitting region is better aligned with the line of sight than the $\gamma$-ray zone, and hence the optical radiation is more beamed
than the $\gamma$-ray emission.}
   \label{fig:gamop_teta}
   \end{figure}

The Large Area Telescope \citep[LAT;][]{atwood2009} onboard the {\it Fermi} satellite has been monitoring the sky in the 0.1--300 GeV energy range since its launch in 2008. 
$\gamma$-ray light curves with monthly, weekly, and three-day binnings are available in the {\it Fermi} LAT Light Curve Repository\footnote{\url{https://fermi.gsfc.nasa.gov/ssc/data/access/lat/LightCurveRepository/}} \citep{abdollahi2023}. 
%This provides calibrated fluxes for more than 1500 sources that were recognized as variable in the fourth Fermi LAT point source catalog \citep[4FGL-DR2;][]{ballet2020}, which covers ten years of sky monitoring by the satellite.
However, the high brightness level of BL Lacertae in the considered period made a sub-daily binning possible, which was desiderable both to investigate the short-term $\gamma$-ray variability, and to compare it with the densely-sampled optical light curve. Therefore, the $\gamma$-ray data were analysed using the FermiTools package installed with Conda\footnote{\url{https://github.com/fermi-lat/Fermitools-conda/}}, with instrument response function P8R3\_V3, Galactic diffuse emission model gll\_iem\_v07, and isotropic background model iso\_P8R3\_ SOURCE\_V3\_v1. 
We considered a region of interest with a radius of $30^\circ$, a maximum zenith angle of $90 ^\circ$, and only ‘source’ class events (evclass=128, evtype=3) for a binned likelihood analysis of the photon data.
The fluxes of the sources within a $10^\circ$ radius were set as free parameters of the model, whereas fluxes of more distant objects were fixed to their mean values according to the 4FGL catalogue. For BL Lacertae (4FGL J2202.7+4216), we adopted a log-parabola spectral model, 
${\rm d} N/{\rm d} E = N_0 \, (E/E_{\rm b})^{-[\alpha+\beta \log(E/E_{\rm b})]}$
, and a break energy of $E_{\rm b} \approx 870.61 \, \rm MeV$. 
%$E_{\rm b} = 870.6099 \, \rm MeV$
During the analysis, the spectral parameters of this source were kept fixed at the values that they assume when integrating over the whole period; that is, $\alpha \approx 2.02, \, \beta \approx 4.43 \times 10^{-2}$.
%$\alpha= 2.021555654, \beta = 4.431249384 × 10^{−2}$.
%, and  test statistic (TS) provided by the maximumlikelihood over the whole period, the fit gives TS= 163558 with an integrated average flux of (8.45773 +/- 0.0237177)10−7 photons cm−2 s−1 . 
The source was considered detected if the test statistic (TS) exceeded 25. 
We repeated the analysis four times with integration bins of 48, 24, 12, and 6 h and then used these data to build a composite light curve with optimised sampling according to the method presented in \citet{raiteri2023b}. 
The resulting $\gamma$-ray light curve is shown in Fig.~\ref{fig:gamop_teta}. The source reached a historical $\gamma$-ray maximum brightness of $(15.2 \pm 0.8)\times 10^{-6} \rm \, ph \, cm^{-2} \, s^{-1}$ (TS=2295) on 2021 April 27 (JD=2459331.6).

\section{Correlation between the $\gamma$-ray and optical emission}
\label{sec:dcf_go}

\begin{figure*}
   \sidecaption
   \includegraphics[width=6.1cm]{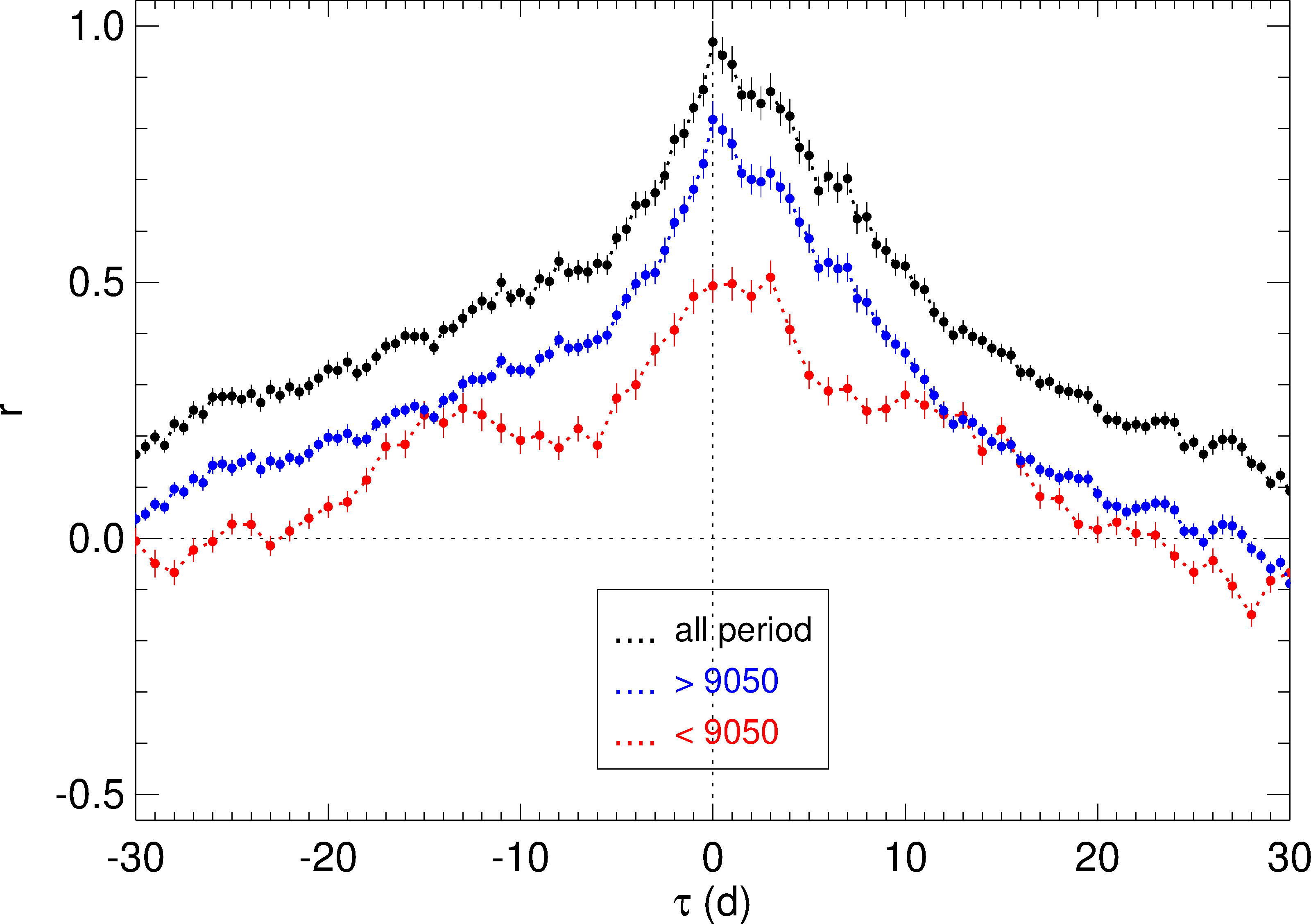}
   \hspace{0.2cm}
   \includegraphics[width=6.1cm]{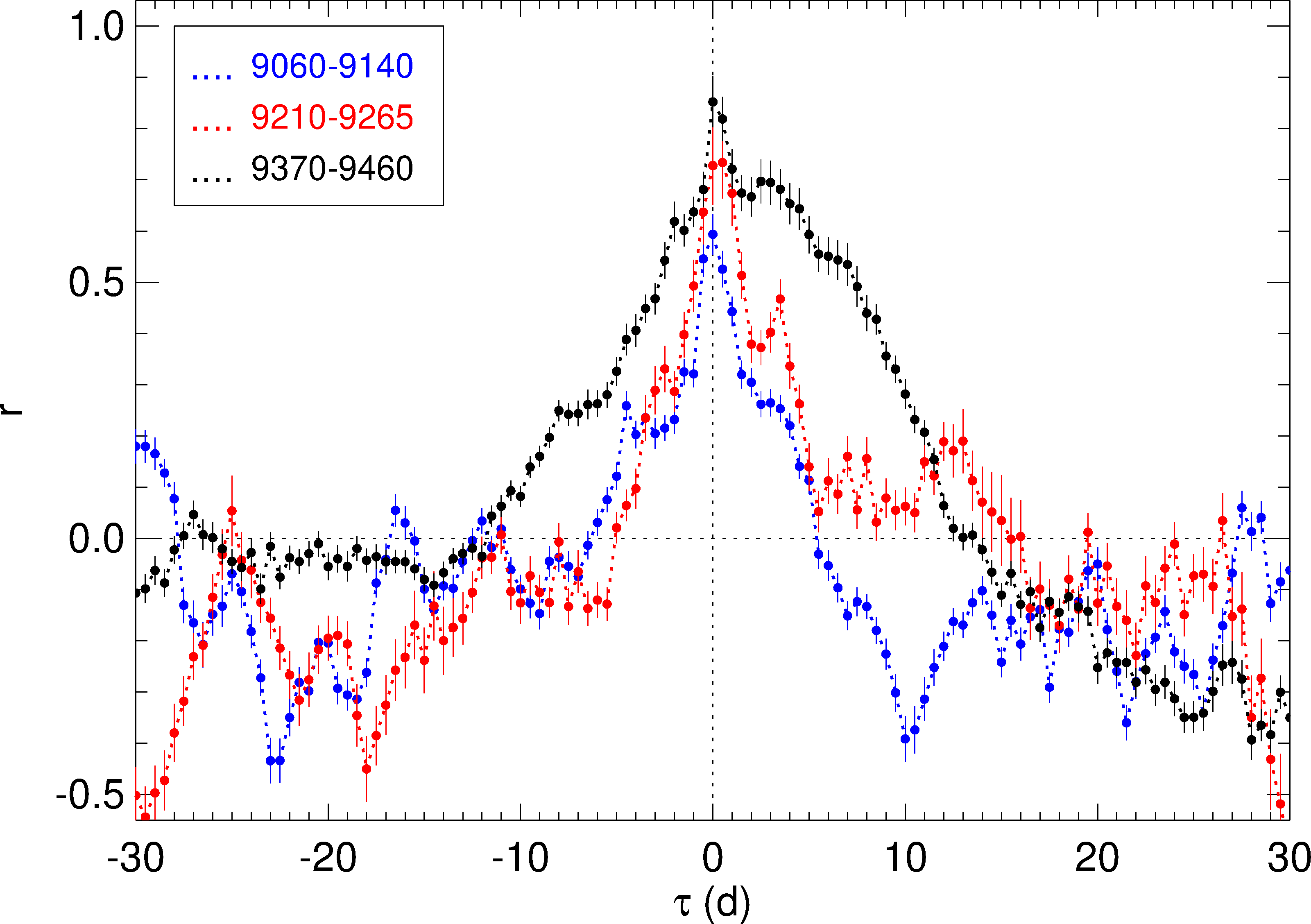}
   \caption{DCF between $\gamma$-ray and optical fluxes for various intervals of time. In all cases, the main peak at $\tau \sim 0 \, \rm d$ indicates correlation with almost no time delay.}
   \label{fig:dcf_go}
\end{figure*}

\begin{figure*}
   \centering
   \includegraphics[width=6cm]{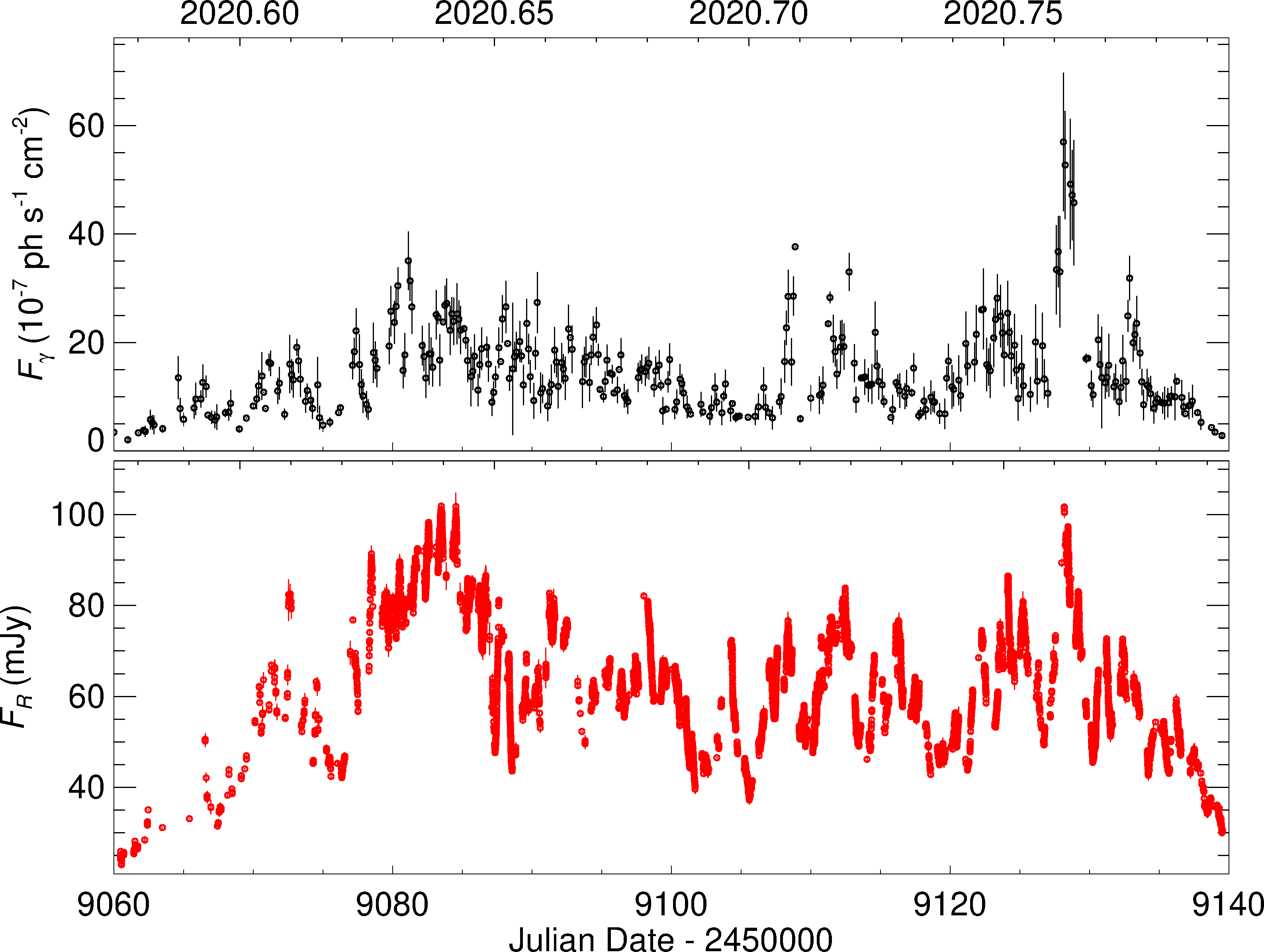}
   \includegraphics[width=5.8cm]{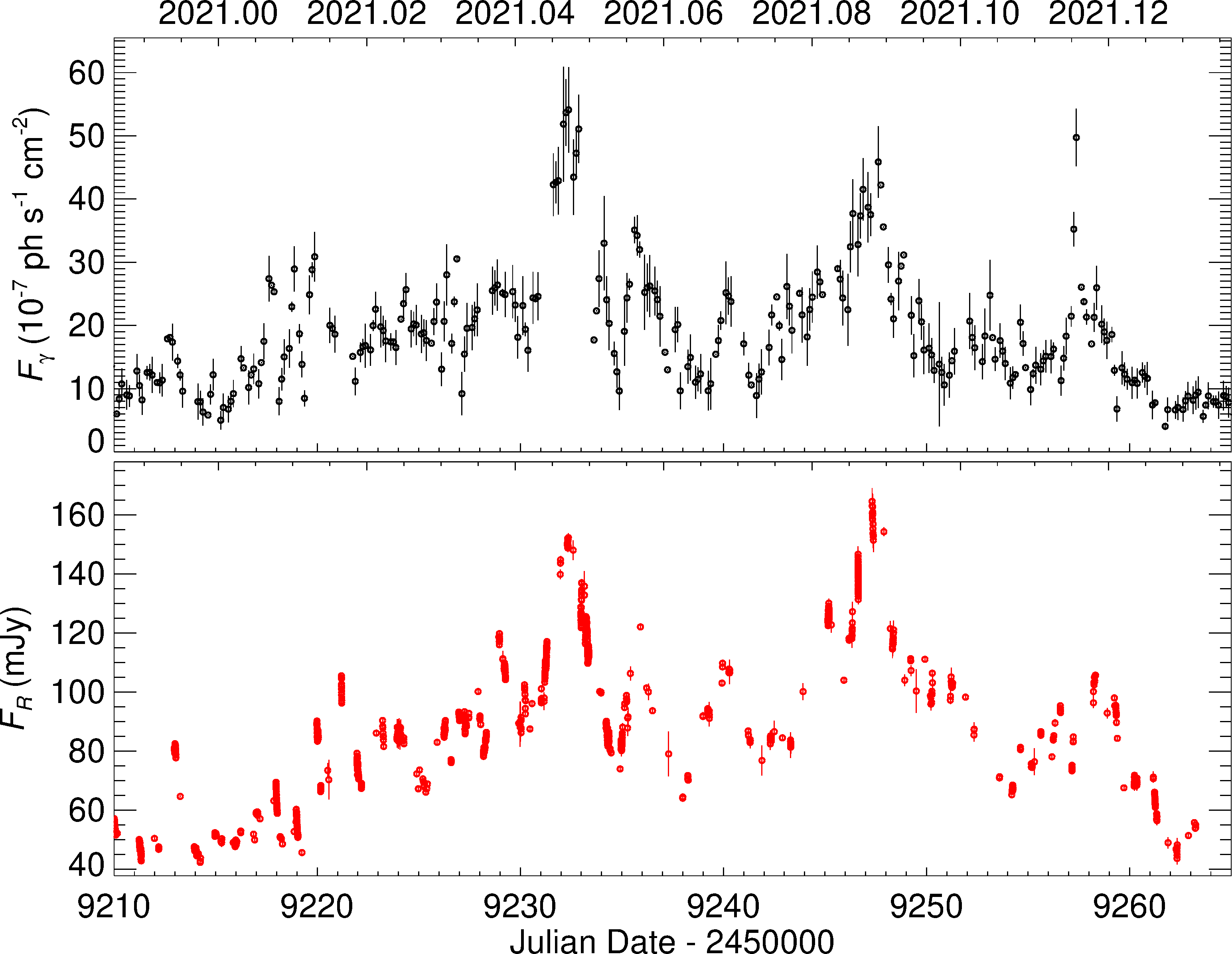}
   \hspace{0.2cm}
   \includegraphics[width=6cm]{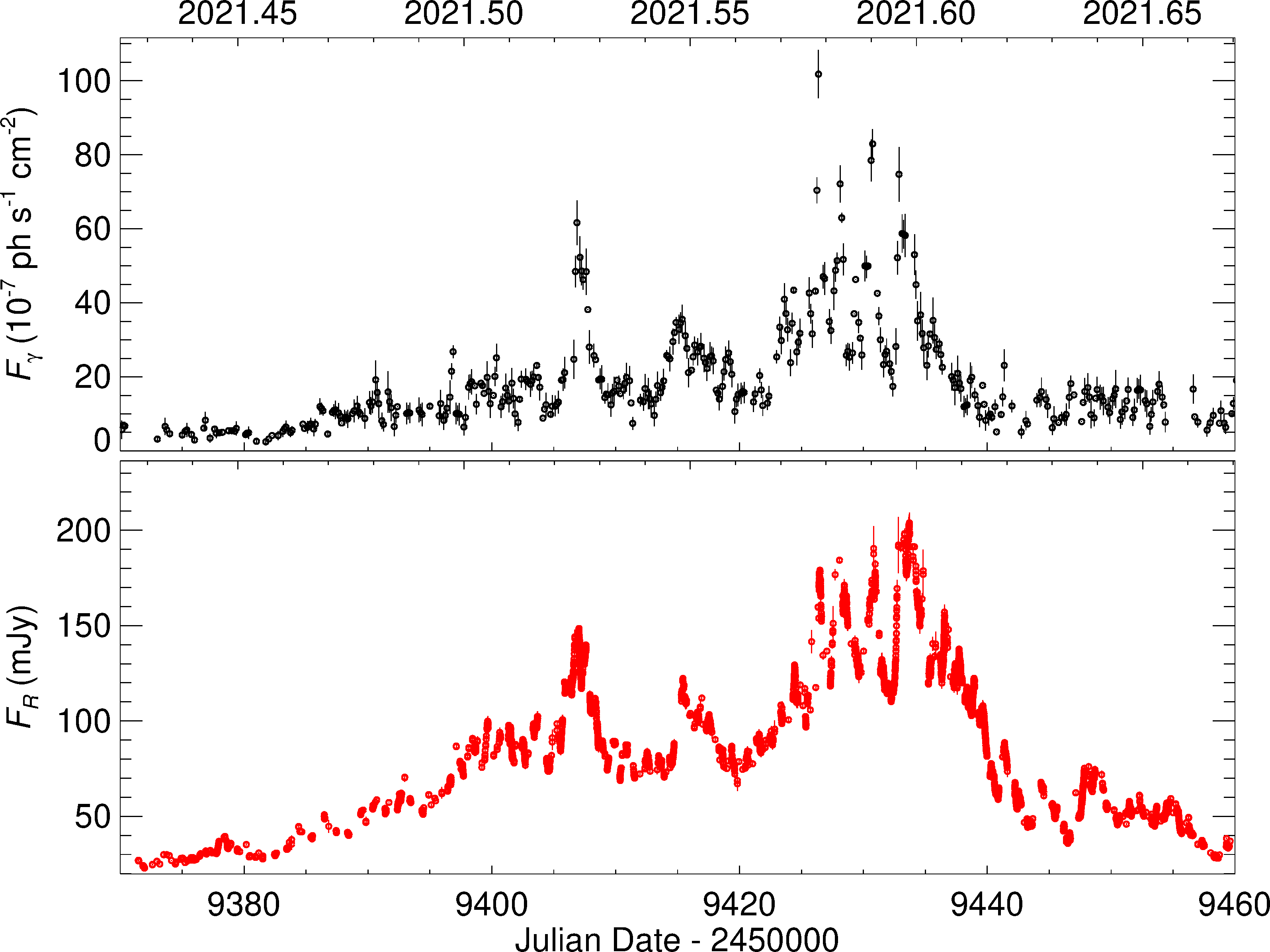}
   \caption{Enlargements of $\gamma$-ray and optical light curves during some of the time intervals considered in Fig.~\ref{fig:dcf_go}.}
   \label{fig:tim}
\end{figure*}

%We also investigated the behaviour of BL Lacertae at high energies in 2019--2022 by means of the $\gamma$-ray data coming from the all-sky continuous monitoring by the {\it Fermi} satellite.
%Figure \ref{fig:gamop_teta} shows the finely sampled $\gamma$-ray light curve that we obtained as explained in Section \ref{gamma}.
The $\gamma$-ray photons are expected to be produced mainly through an inverse-Compton scattering of soft photons off the same relativistic electrons that emit the optical radiation.  The origin of these soft seed photons is still debated.
%Therefore, the $\gamma$-ray radiation should correlate with the optical one, and they should come from roughly the same jet region. This implies that the viewing angles derived from the $\gamma$-ray and optical data should approximately correspond, and the $\gamma$-ray and optical fluxes should be affected by the same Doppler beaming.
We analysed the correlation between the $\gamma$-ray and optical fluxes with the DCF for various time intervals. 
The results are shown in Fig.~\ref{fig:dcf_go}.
Enlargements of the $\gamma$-ray and optical light curves in some of these periods are displayed in Fig.~\ref{fig:tim}.

All DCFs show a main peak at a time lag of $\tau \sim 0 \, \rm d$, indicating almost simultaneous variations in the $\gamma$-ray and optical fluxes. However, the strength of the correlation ranges from $\sim 0.5$ (fair correlation) to $\sim 1$ (good correlation). In particular, the lowest value is found for the 2019--2020 period, when the source was fainter and the sampling less dense than in the subsequent flaring states.
%Moreover, in most cases the central peak appears somewhat skewed towards (small) positive time lags, indicating a delay of the optical flux variations after the $\gamma$-ray ones.
%This asymmetry in the DCF is likely due to the 2021 major optical outburst lasting longer than its $\gamma$-ray counterpart because of the small misalignment of the corresponding emitting regions. Indeed, the DCF between the $\gamma$-ray and optical viewing angles in the period JD=2459370--2459460 shows a delay of $\theta$ optical with respect to $\theta$ $\gamma$ of about two days (see Fig.~\ref{fig:dcf_pete_teta_go}).  
%This is likely due to fast cooling of the upscattering electrons, which implies that inverse-Compton scattering provides softer and softer $\gamma$-ray photons that fall outside the range of energy of the $\gamma$-ray photons we are analysing.

\section{Twisting-jet model applied to the $\gamma$-ray emission}
\label{sec:go}
We followed a procedure similar to that applied to the optical and radio data to derive the behaviour in time of the viewing angle from the $\gamma$-ray fluxes.
To obtain the behaviour in time of $\delta$ and $\theta$ corresponding to the $\gamma$-ray light curve, we first derived the long-term trend in the $\gamma$-ray band by means of a cubic spline interpolation through the binned $\gamma$-ray fluxes in the same optical variable time bins. This is shown in Fig.~\ref{fig:gamop_teta}.
Then, we rescaled the spline as
$F_{\rm resc}=a \, F^b$ \citep[see also][]{raiteri2011}, where the coefficients $a$ and $b$ were fixed by constraining the $\gamma$ spline to match the same minimum and maximum values as the optical one (see Fig.~\ref{fig:gamop_teta}).
Finally, we obtained $\delta$ and $\theta$ from $F_{\rm resc}$ in the same way as in the optical case.

The last panel of Fig.~\ref{fig:gamop_teta} shows the comparison between the optical and $\gamma$-ray viewing angles. In general, they are in good agreement, with a mean difference of $0.06^\circ$.  The main discrepancies are limited to some short periods and depend, at least in part, on the different sampling of the two bands.
%NB Non usiamo gli errori nei fit perche` sono misleading, in quanto nella curva luce composita i punti a 6h hanno errori maggiori di quelli a 12 h, che hanno errori maggiori di quelli a 24 h.
Moreover, small space displacements between the photocentres of the optical and $\gamma$-ray emissions may occasionally occur, which would imply a difference in beaming and thus explain the presence of sterile optical flares (i.e. without $\gamma$-ray counterpart) or orphan $\gamma$-ray flares (i.e. without optical counterpart) that are also observed in other blazars.
However, other interpretations for uncorrelated optical and $\gamma$-ray flares are possible, involving hadronic processes \citep[e.g.][]{dejaeger2023} or other mechanisms \citep[e.g.][]{macdonald2017,sobacchi2021,wang2022}.

\begin{figure}
   \resizebox{\hsize}{!}{\includegraphics{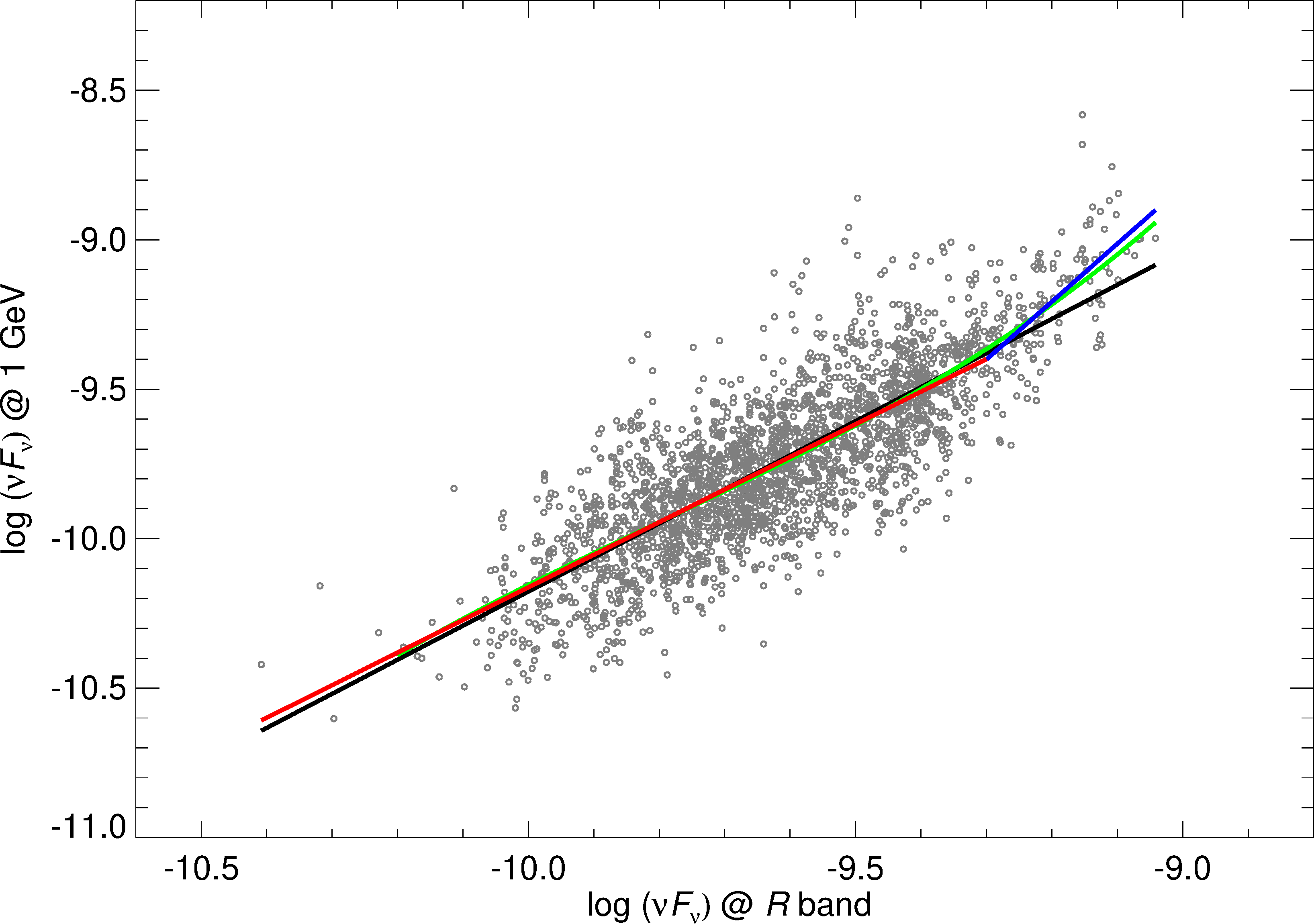}}
      \caption{$\gamma$-ray fluxes at 1 GeV versus optical fluxes in $R$ band. Each data point of $\gamma$-ray light curve has been paired with the average of the optical points acquired within 6~h. Solid lines represent linear regressions to the whole dataset (black), to the data points with optical $\log (\nu \, F_\nu)$ less than -9.3 (red), and to the data points with optical $\log (\nu \, F_\nu)$ greater than -9.3 (blue).
      The green line represents a cubic regression.
              }
         \label{fig:go}
   \end{figure}

In summary, the comparison between the optical and $\gamma$-ray light curves suggests that the flux variations in the two bands are generally well correlated and that both emissions originate in the same jet region and are therefore subject to the same Doppler beaming.
%The relationship between the $\gamma$ and optical fluxes is also determined by the nature of the soft photons that are upscattered. 
This is in agreement with a synchrotron self-Compton (SSC) origin of the $\gamma$-ray photons, according to which the soft seed photons for the inverse-Compton process are the same optical photons produced by the synchrotron process. 
We further investigated this issue through a direct comparison of 
%Therefore, we expect a blend of linear (geometric) and quadratic (SSC) dependence of the $\gamma$-ray fluxes on the optical ones.
%Figure~\ref{fig:go} shows the correlation between 
the $\gamma$-ray fluxes [$\log (\nu \, F_\nu)$ at $\log \nu=23.38$; i.e. 1 GeV] with the optical fluxes  [$\log (\nu \, F_\nu)$ at $\log \nu=14.67$; i.e. the effective frequency of the $R$ band]. The result is shown in Fig.~\ref{fig:go}.
Each data point in the $\gamma$-ray light curve has been coupled with the average of the optical points acquired within 6~h, leading to 2334 pairs.

A linear regression on the whole dataset gives a slope of $k=1.14 \pm 0.02$, which corresponds to the index of the power law $F_\gamma \propto F_{R}^k$. 
However, we expect two different mechanisms to be at work.
%, i.e., the geometrical effect on the long-term trend and intrinsic processes on the short-term variations. 
The geometric effect on the long-term trend alone would yield a linear regression with an $\sim 1$ slope, because both optical and $\gamma$-ray fluxes are affected by the same Doppler boosting.
In contrast, the energetic processes in the jet that are responsible for the short-term variability, together with an SSC nature of the $\gamma$-ray radiation, would lead to an $\sim 2$ slope. 
%Indeed SSC implies that the production of $\gamma$-ray photons depends on both the relativistic electrons emitting the optical photons, and on the optical photons themselves, which act as seed photons for the inverse-Compton scattering. This leads to a quadratic dependence of the $\gamma$-ray flux on the optical flux.
The 1.14 slope we found suggests that the dominant mechanism is the geometric one.
However, if we fit the data with a cubic regression, the curve deviates from the linear regression in the bright end, starting from about $\log (\nu  \, F_\nu) = -9.3$ in the $R$ band (i.e. $\sim 107$ mJy), indicating a steeper slope.
 Indeed, if we restrict the linear regression to the 141 points of this bright end,
 %with $\log (\nu  \, F_\nu)> -9.3$ in the $R$ band 
 we obtain a slope of $1.94 \pm 0.20$, consistent with 2, while for lower optical fluxes the slope becomes $1.09 \pm 0.02$, which is very close to 1. This is consistent with our scenario, because the highest optical fluxes correspond to very rapid flares of intrinsic nature, and therefore the signature of the SSC relationship between the optical and $\gamma$-ray fluxes emerges in the slope. Actually, the dispersion of the data points around the slope's $\sim 1$ regression line is caused by the short-term variations (with a slope of $\sim 2$) superimposed on the long-term variability of geometric origin.

\section{Testing the geometric model}
\label{sec:tests}
The reliability of our geometric interpretation of the blazar's long-term variability can be checked through a deeper analysis of the optical light curve.
The model implies that we should find some signatures in the observed source behaviour.
%some consequences that must find an observational assessment. 
However, the contrary is not necessarily true; as \citet{scargle2020} pointed out, one should avoid inferring physical properties from statistical properties of observed time series.

\subsection{Flux distribution}
\label{distri}

\begin{figure*}
   \sidecaption
   \includegraphics[width=6.cm]{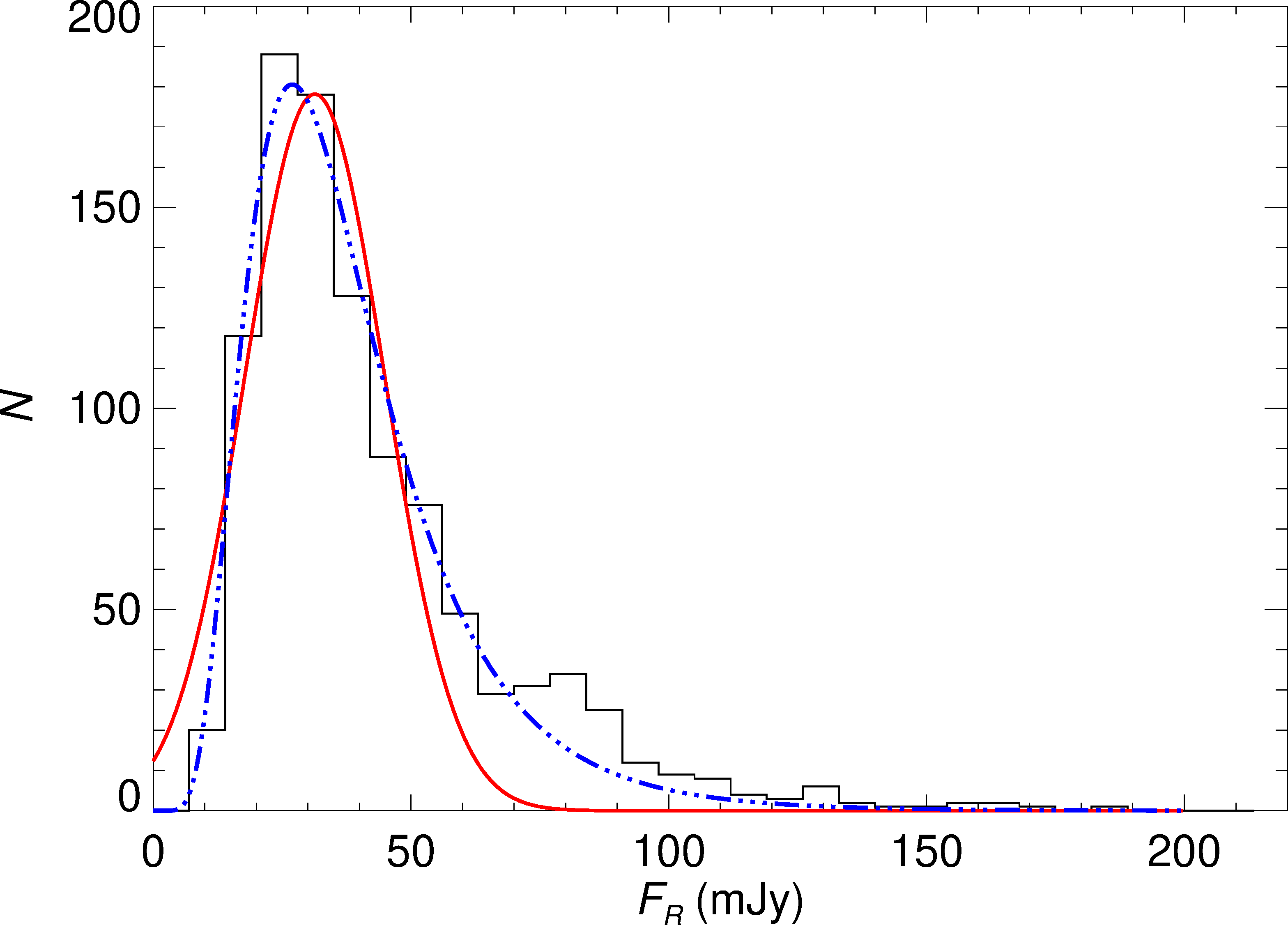}
   \hspace{0.2cm}
   \includegraphics[width=6.1cm]{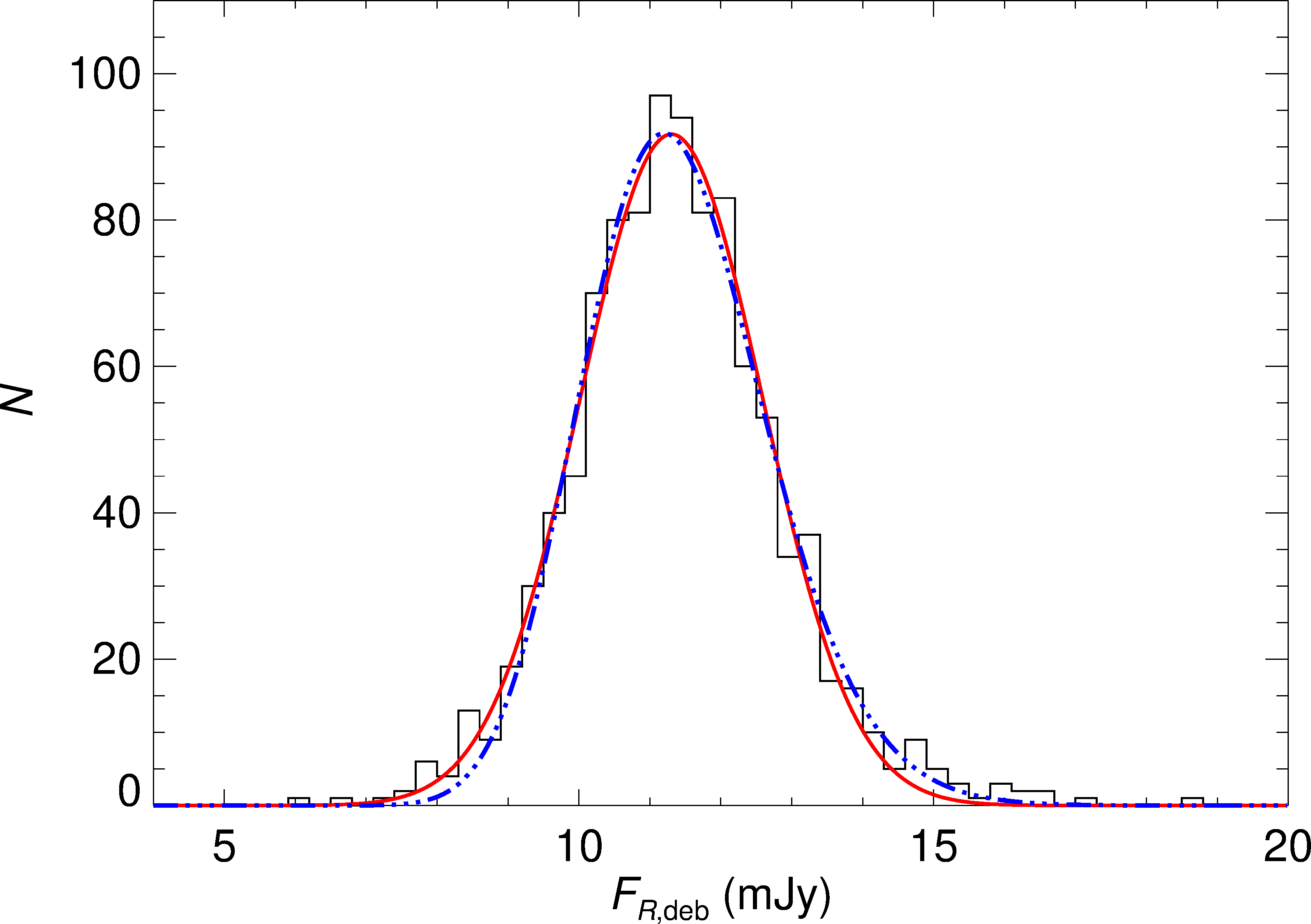}
      \caption{Distribution of the optical flux densities. Left: Flux densities corrected for Galactic extinction and host galaxy contribution. Right: The same flux densities after deboosting, that is after correcting for the variable Doppler beaming. Red continuous lines and blue dotted-dashed lines represent Gaussian and Lognormal fits, respectively. 
              }
         \label{fig:distri}
   \end{figure*}

One of the signatures that we can investigate is the flux distribution.
The flux distribution has been analysed in several papers in order to find hints on the underlying physical mechanism.
%Several papers have discussed the possibility to infer information on the variability mechanisms by analysing the flux distribution of the source. 
This was mainly done for the X-ray light curves of X-ray binaries and AGNs \citep[e.g.][]{uttley2005} and then extended to blazars \citep[e.g.][]{sinha2018}.
In general, it was suggested that Gaussian distributions can be the result of additive processes, while multiplicative processes would led to log-normal distributions.
We built the distribution of the optical flux densities for the period considered in this paper. To mitigate the effect of dense intra-day sampling, we first binned the $R$-band light curve every day and then in one-week bins. The resulting flux distribution, which is displayed in Fig.~\ref{fig:distri}, is clearly skewed towards the lowest fluxes, and is better fitted by a log-normal (reduced $\chi^2=48.7$) than a Gaussian (reduced $\chi^2=308$). 

When we consider the deboosted fluxes (see Fig.~\ref{fig:radop_teta}), we obtain a more symmetric distribution, which is best fitted by a Gaussian (reduced $\chi^2=11.1$) than a log-normal (reduced $\chi^2=12.2$).
These results are consistent with the twisting-jet model. The observed flux densities depend on the Doppler beaming, which increases when the emitting region becomes better aligned with the line of sight. 
If the jet wiggles randomly, this occurs not very often, and indeed major outbursts are rare events. Therefore, the highest flux values build a tail in the flux distribution. 
The deboosted flux densities instead represent the jet emission in the case of a constant Doppler factor. They differ by only a scale factor from the intrinsic flux densities, whose variability is likely due to energetic processes occurring in a region of the jet composed by several sub-zones; the contributions of these sub-zones to the total flux are additive. We note that jet sub-zones seem to be necessary to explain the very fast variability observed in the optical band and at other frequencies \citep[e.g.][]{raiteri2023b,begelman2008,shukla2020}.

\subsection{Signatures of time-dependent Doppler beaming}
\label{short}
\begin{figure*}
   \sidecaption
   \includegraphics[width=6.cm]{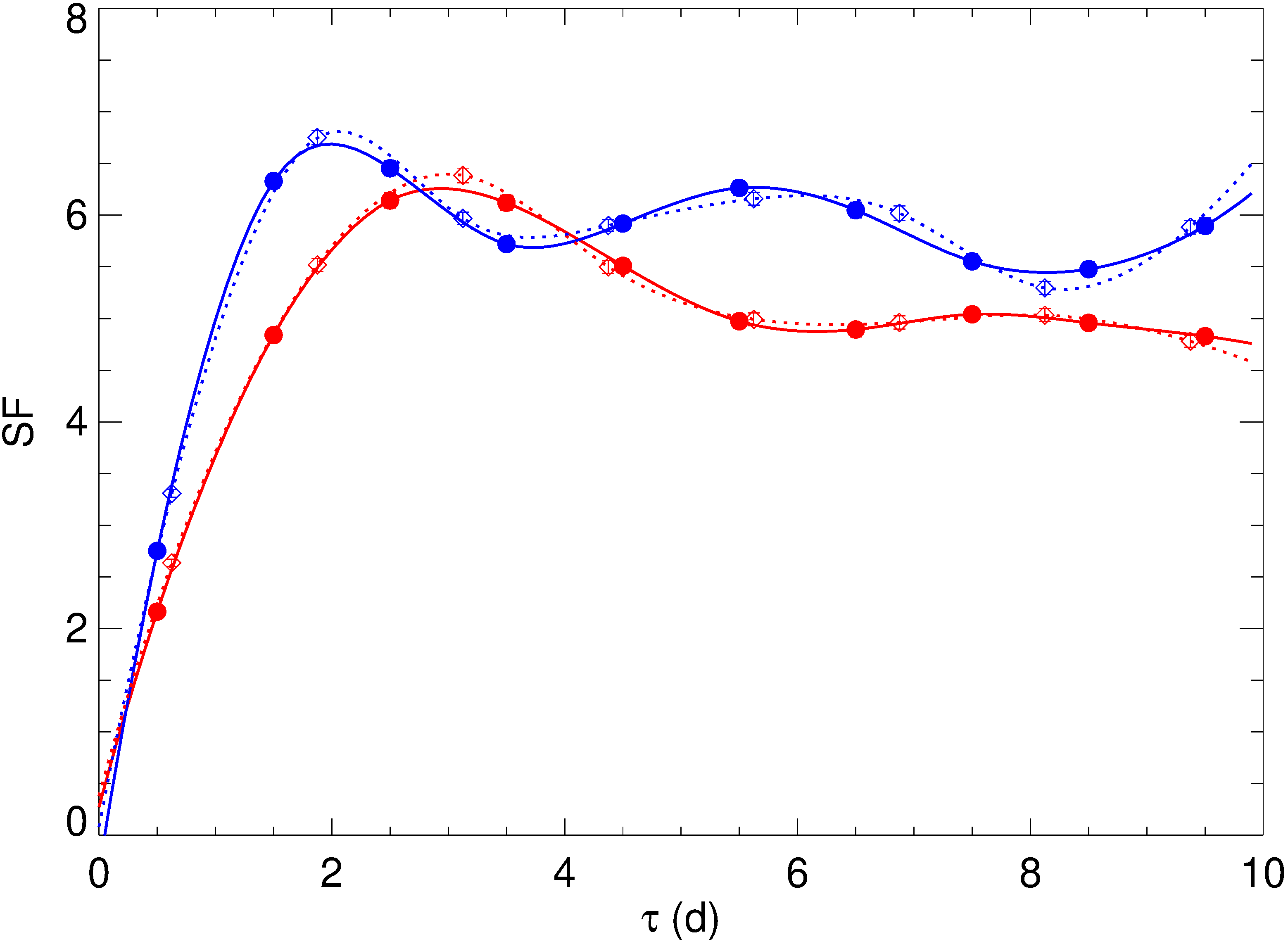}
   \hspace{0.2cm}
   \includegraphics[width=6.2cm]{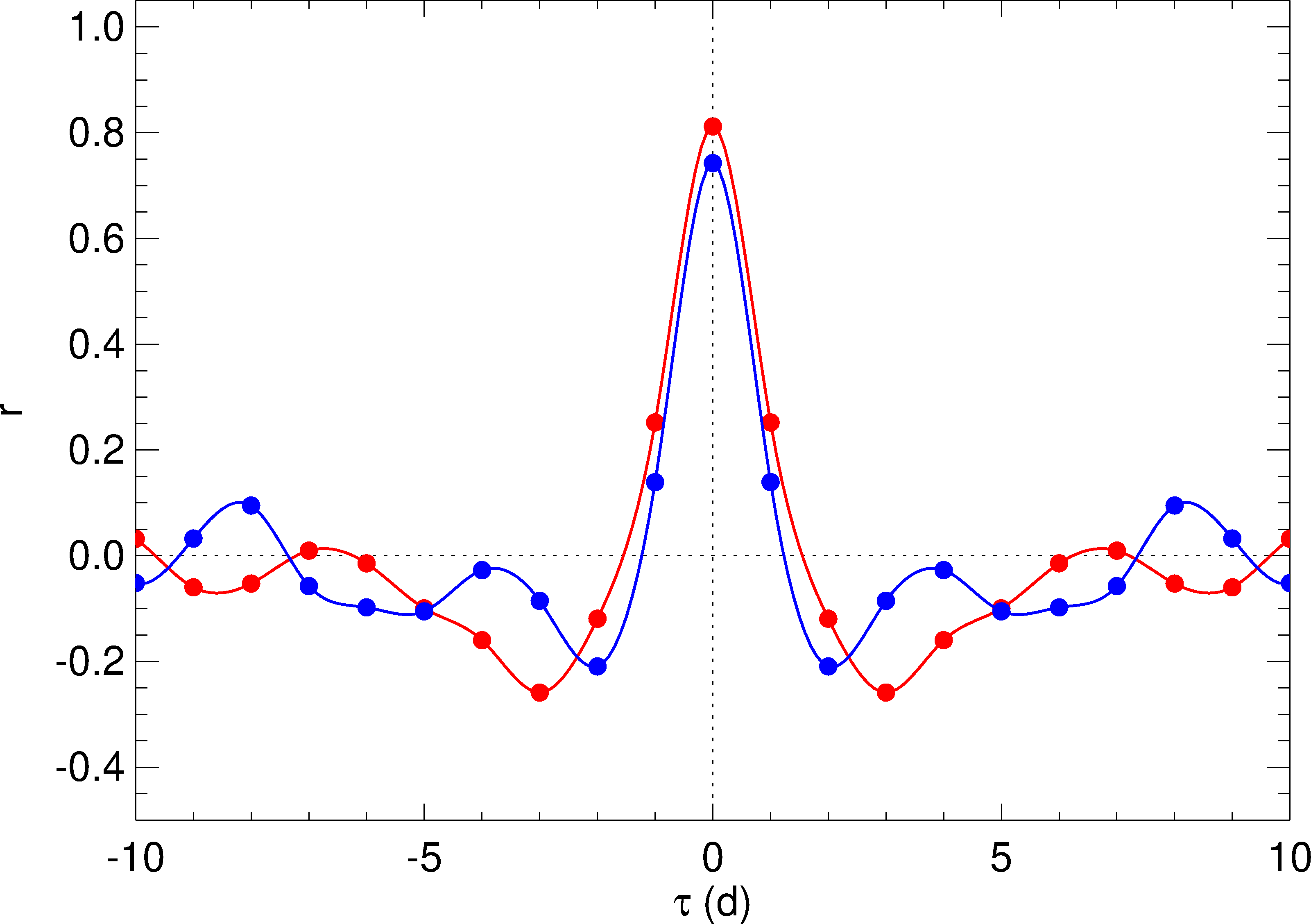}
      \caption{Results of the time-series analysis. Left: SF on deboosted optical flux densities in low (red) and high (blue) states. Dots refer to SF sampling every 24 h, and empty diamonds refer to sampling every 30 h. Right: ACF on deboosted optical flux densities in low (red) and high (blue) states. 
      In both panels, lines represent cubic spline interpolations.
              }
         \label{fig:timescales}
   \end{figure*}

\begin{figure}
   \includegraphics[width=\hsize]{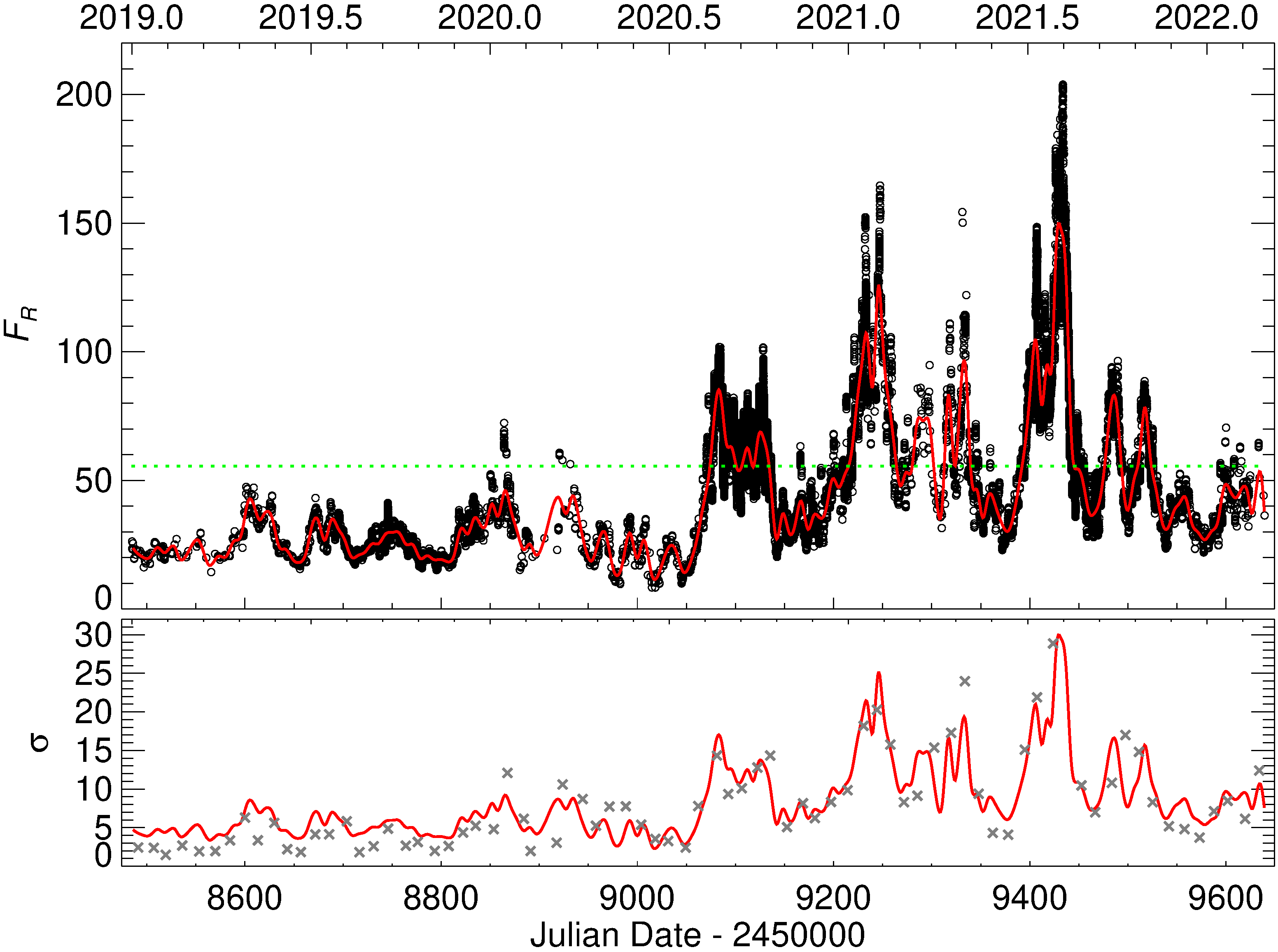}
      \caption{Increase of the flux-variation amplitude with brightness. Top: Optical flux densities (black circles) and their long-term trend (red solid line). The green horizontal dotted line indicates the level of flux corresponding to the median value of the Doppler factor, $\delta_{\rm med} = 14.96$. Bottom: Standard deviations of optical flux densities in bins of 15 d (grey crosses) compared with the long-term trend (red line) rescaled by a factor of five.}
         \label{fig:doppler}
   \end{figure}

Doppler beaming produces a 
shortening of the observed variability timescales with respect to those in the rest frame $\Delta t_{\rm obs}= \Delta t_{\rm rest}/\delta$.
If we assume that the long-term variations observed in a light curve are due to changes in the Doppler factor, we should find that variability timescales are shorter when the flux is higher.
To check this effect, we calculated the structure function \citep[SF;][]{simonetti1985} and the ACF on the deboosted optical flux densities in low and high states. 
Low and high states are defined as those characterised by Doppler factors lower and higher than their median value, respectively (see Fig.~\ref{fig:doppler}). 
Choosing the median value of $\delta$, which corresponds to 14.96, implies that we have the same number of points in low and high states. 
%14.955
The results are shown in Fig.~\ref{fig:timescales}.
In all cases, the data were first averaged over 1 h time bins to mitigate the effect of dense intra-night sampling.
Timescales can be derived as the lags $\tau$ where the SF reaches peaks and where the ACF shows minima.
The SF corresponding to the low states has the first peak at $\tau =2.9$--3.0 d, depending on the SF bin. In the case of high states, the first peak corresponds to $\tau =2.0$--2.1 d.
The ratios between the shortest timescale in low and high states is $\sim 1.4$.
The same result is obtained with the ACF, where the ratio between the timescales of the first minimum in low and high states is again 1.4.
This compares fairly well with the ratio between the maximum $\delta$ in high and low states, which is $\sim 1.3$.

\begin{figure}
   \includegraphics[width=\hsize]{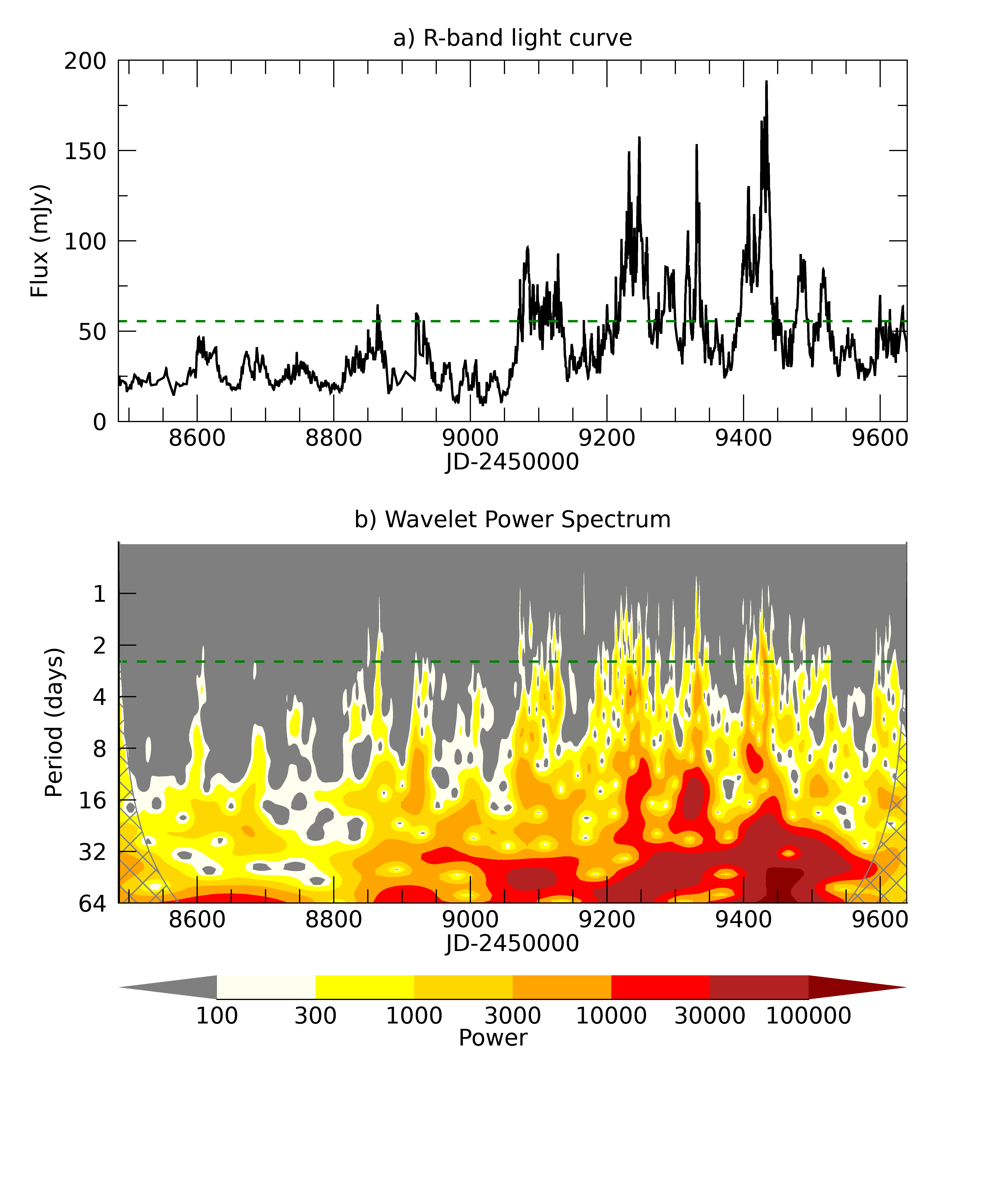}
   %\vspace{-1.5cm}
      \caption{Results of the wavelet analysis. a) Optical light curve of BL Lacertae in 2019--2022. The green dashed line represents the flux level corresponding to the median value of the Doppler factor, $\delta_{\rm med} = 14.96$. b) Wavelet spectrum for optical light curve. The green dashed line marks a timescale of 2.5 d, which is between those found in high- and low-brightness states with the SF and ACF (see Fig.~\ref{fig:timescales}). }
         \label{fig:wavelet}
   \end{figure}

Support to the timescale decrease in bright states discussed above come from the wavelet analysis \citep{torrence1998}. The wavelet spectrum is shown in Fig.~\ref{fig:wavelet}. It is easy to see that timescales shorter than 2.5 d are present when the source flux undergoes a Doppler beaming greater than its median value.

   Moreover, if the long-term behaviour is due to variations in the Doppler factor, we expect a flux-variability amplitude that increases with brightness, since the flux amplitude has the same dependence on $\delta$ as the flux density; that is, $\Delta F_\nu \propto \delta^4$ in the optical band \citep[see Eq.~\ref{eq:flux} and][]{raiteri2017_nature}.
   To check this effect, we calculated the standard deviations of the optical flux densities over bins of 15 d. The results are shown in Fig.~\ref{fig:doppler} and confirm that the flux variability amplitude is larger when the source is brighter. 
   Indeed, the comparison between the standard deviations and the long-term trend shows a good agreement, the small discrepancies being essentially due to inhomogeneous sampling.

\section{Summary and discussion}
\label{sec:end}

In this work, we analysed optical and radio data of BL Lacertae acquired by the WEBT and other facilities in 2019--2022 and $\gamma$-ray data from the {\it Fermi} satellite in the same period.
Because of the source brightness, the corresponding light curves are extremely well sampled and allowed us to study the multi-wavelength behaviour in detail.

We applied the geometric jet model by \citet{raiteri2017_nature} and derived the twisting motion of the optical-, radio-, and $\gamma$-ray-emitting regions.
The cross-correlation between the optical and radio viewing angles shows multiple signals, the strongest one indicating that the radio lags behind the optical by $\sim 120 \, \rm d$. We interpret these signals as being due to two interlaced twisting jet filaments, in each of which the optical-emitting region lies upstream with respect to the radio one. Due to jet rotation, the emitting regions of each filament recurrently achieve the minimum viewing angle, and, consequently, the maximum Doppler beaming, every $\sim 200$ d. 

The comparison between the $\gamma$-ray behaviour and the optical one shows that the two emissions are strongly correlated and mostly co-spatial, which is in agreement with the above geometric scenario and confirms an SSC origin of the $\gamma$-ray radiation.
The predictions of the model on the flux distributions and on the changes in the variability timescales and amplitudes with brightness are verified.
%In conclusion, we propose that the long-term flux variability observed in the multiwavelength light curves of BL Lacertae is caused by variations in the Doppler beaming due to changes in the viewing angles of the emitting zones in a wiggling jet composed of a pair of interlaced helical filaments.
%Previous results showing the presence of obtained for other blazars suggest that this model can account for blazar variability in general.
%The analysis of the variability behaviour in other blazars indicated a similar twisting structure, so that the model here inferred for BL Lacertae is expected to hold for the blazar class as a whole. 

The wiggling filamentary model finds support both on observational and theoretical grounds.

Extragalactic jets are expected to rotate around their axis because of the rotation of the accretion disc and supermassive black hole from which they are launched, but an observational confirmation is difficult to obtain. Hints in favour of rotation were detected by \citet{mertens2016} in the jet of M~87 and by \citet{fuentes2023} in the jet of 3C~279.
Numerical simulations of rotating jets were performed, giving insight into the effect of rotation on plasma instabilities (e.g. \citealt{bodo2019} and references therein). 
%found that rotation has a stabilisation effect on the jet against some plasma instabilities, but can produce others \citep[e.g][and references therein]{bodo2019}.
%\citep{bodo1996,nakamura2004,carey2009,bodo2019}. 
%According to \citet{fendt2011}, MHD shocks in a helical magnetic field can produce jet rotation at a speed of 0.1--1\% of the jet bulk velocity.
In our scenario, the emitting regions along each filament recurrently approach the line of sight with a timescale for completing a whole turn of $\sim 200 \, \rm d$. This implies an angular velocity of $\omega \approx 3.6 \times 10^{-7} \rm  \, s^{-1}$, which is of the order  (2.5 times lower) of the value estimated by \citet{mertens2016} for the inner region of the M~87 jet they analysed.
%vedi anche \citet{sobyanin2017} f

Regarding the twisting structure, radio images of extragalactic jets have often shown an oscillating pattern \citep[see e.g.][]{owen1980,giroletti2004,perucho2012,fromm2013,casadio2015,britzen2017,britzen2018,lister2021,issaoun2022,zhao2022,pushkarev2023}. Numerical simulations have found that plasma instabilities that develop in the jet can give rise to twisting structures \citep[e.g.][]{nakamura2001,hardee2003,moll2008,mignone2010}. 
%The simulation of orbiting or precessing nozzles also led to curved jet motion \citep{fendt2022}.
Orbital motion in a binary black hole system or jet precession can also produce bending and twisting structures \citep{begelman1980,liska2018,fendt2022}.

Finally, a multiple filamentary structure of extragalactic jets has been observed through high-resolution radio images in M~87 \citep[e.g.][and references therein]{owen1989,nikonov2023} 
%and NGC~315 \cite{worrall2007}, 
and in the FSRQs 3C~273 \citep{lobanov2001} and 3C~279 \citep{fuentes2023}. 
Theoretical arguments and numerical simulations support the presence of such filaments.
Winding helical filaments emerge as exact solutions for the magnetohydrodynamic (MHD) equilibrium of astrophysical jets \citep{villata1994,villata1995}.
\citet{lesch1998} proposed that a filamentary structure can naturally arise in MHD flows and that magnetic reconnection in such a configuration can provide an efficient mechanism for particle acceleration.
This scenario was further investigated in \citet{wiechen1998} by means of MHD simulations, showing that perturbations such as Kelvin-Helmholtz instabilities can lead to filamentary structures.
The threads observed in the M~87 jet have often been ascribed to Kelvin-Helmholtz instabilities \citep[e.g.][]{lobanov2003,hardee2011,nikonov2023}.
In the MHD simulations by \citet{mizuno2012}, the growth of current-driven kink instabilities in a jet with a helical magnetic field led to twisted magnetic filaments.

%Moreover, a filamentary structure implies smaller emitting regions, providing an explanation for the very fast variability observed in the multiwavelength light curves of several blazars, including BL Lacertae \citep[e.g.][and references therein]{fuentes2023,raiteri2023b,raiteri2023a}.

%All of the above findings suggest that extragalactic jets are formed by twisting filamentary structures in rotation. 
%In previous works the analysis of well-sampled multifrequency light curves of some blazars led us to proposed a twisting jet model. In this paper we applied this model to the radio, optical, and gamma-ray behaviour of BL Lacertae in 2019--2022 and further suggest that the twisting jet is made up of filaments.

In the light of all the above results, we believe that the wiggling filamentary jet model here proposed for BL Lacertae can provide a valid description of extragalactic jets. The applicability of this model to other blazars can be verified in the near future through the synergy between the multi-wavelength study of the blazar emission and the very high resolution of the radio images achievable through space-VLBI observations.

   \section*{Data Availability}
    Data acquired by the WEBT Collaboration are stored in the WEBT archive and are available upon request to the WEBT President Massimo Villata (\href{mailto:massimo.villata@inaf.it}{massimo.villata@inaf.it}). 
    {\it Fermi}/LAT data can be downloaded from the National Aeronautics and Space Administration (NASA) site (\url{https://fermi.gsfc.nasa.gov/ssc/data/access/}); the $\gamma$-ray light curve published here can be obtained from the authors upon request.

\begin{acknowledgements}
We thank Eduardo Ros (Max-Planck-Institut f\"ur Radioastronomie, Bonn, Germany) for insightful comments on the original version of the manuscript.
We are grateful to 
M. L. Lister,
to
E. Egron,
D. Perrodin, and
M. Pili (INAF, Osservatorio Astronomico di Cagliari, Italy),
to D. Lane (Burke-Gaffney and Abbey Ridge observatories, Canada),
to
N. A. Nizhelsky,
G. V. Zhekanis,
P. G. Tsybulev, and 
A. K. Erkenov (Special Astrophysical Observatory of RAS, Russia),
to
W. J. Hou,
C. S. Lin, and 
H. Y. Hsiao (National Central University, Taiwan),
and to
L. Tilke and
C. Singh (Connecticut College, USA) for support to the observations.
We are indebted to Simona Villata and Giorgio Mogli for their help with the jet model graphics.
This research has made use of NASA’s Astrophysics Data System Bibliographic Services and of the NASA/IPAC Extragalactic Database, which is funded by the National Aeronautics and Space Administration and operated by the California Institute of Technology.
The INAF-OATo team acknowledges financial support from the INAF Fundamental Research Funding Call 2023. 
This research has made use of data from the OVRO 40-m monitoring program \citep{richards2011}, supported by private funding from the California Insitute of Technology and the Max Planck Institute for Radio Astronomy, and by NASA grants NNX08AW31G, NNX11A043G, and NNX14AQ89G and NSF grants AST-0808050 and AST-1109911.
This research has made use of data from the MOJAVE database that is maintained by the MOJAVE team \cite{lister2018}. 
The Medicina radio telescope is funded by the Ministry of University and Research (MUR) and is operated as National Facility by the National Institute for Astrophysics (INAF).
The Sardinia Radio Telescope is funded by the Ministry of University and Research (MUR), Italian Space Agency (ASI), and the Autonomous Region of Sardinia (RAS) and is operated as National Facility by the
National Institute for Astrophysics (INAF).
The research at Boston University was supported by National Science Foundation grant AST-2108622 and NASA Fermi Guest Investigator grants 80NSSC22K1571 and 80NSSC23K1507.
This study used observations conducted with the 1.8 m Perkins Telescope Observatory (PTO) in Arizona (USA), which is owned and operated by Boston University.
The IAA-CSIC group acknowledges financial support from the grant CEX2021-001131-S funded by MCIN/AEI/10.13039/501100011033 to the Instituto de Astrof\'isica de Andaluc\'ia-CSIC. 
The IAA-CSIC activities were also supported by MICIN through grants PID2019-107847RB-C44 and PID2022-139117NB-C44. 
The POLAMI observations were carried out at the IRAM 30m Telescope. IRAM is supported by INSU/CNRS (France), MPG (Germany) and IGN (Spain). 
Some of the data are based on observations collected at the Centro Astron\'omico Hispano en Andaluc\'ia (CAHA), operated jointly by Junta de Andaluc\'ia and IAA-CSIC. 
Some of the data are based on observations collected at the Observatorio de Sierra Nevada, owned and operated by the IAA-CSIC.
This paper is partly based on observations made with the IAC-80 telescope operated on the island of Tenerife by the Instituto de Astrof\'isica de Canarias in the Spanish Observatorio del Teide and on observations made with the LCOGT 0.4 m telescope network, one of whose nodes is located in the Spanish Observatorio del Teide.
NRIAG team acknowledges financial support from the Egyptian Science, Technology \& Innovation Funding Authority (STDF) under grant number 45779.
We acknowledge support by Bulgarian National Science Fund under grant DN18-10/2017 and Bulgarian National Roadmap for Research Infrastructure Project D01-326/04.12.2023 of the  Ministry of Education and Science of the Republic of Bulgaria.
The R-band photometric data from the University of Athens Observatory (UOAO) were obtained in the frame of {\it BOSS Project}, after utilizing the robotic and remotely controlled instruments at the University of Athens \citep{gazeas2016}.
This research was partially supported by the Bulgarian National Science Fund of the Ministry of Education and Science under grants KP-06-H38/4 (2019) and KP-06-H68/4 (2022). The Skinakas Observatory is a collaborative project of the University of Crete, the Foundation for Research and Technology -- Hellas, and the Max-Planck-Institut f\"ur Extraterrestrische Physik.
The work by Y.V.S., T.V.M., Y.A.K., A.V.P. is supported by the Ministry of Science and Higher Education of the Russian Federation under the contract 075-15-2024-541.
M.D.J. thanks the Brigham Young University Department of Physics and Astronomy for continued support of the ongoing extragalactic monitoring program at the West Mountain Observatory.
This work is partly based on observations carried out
at the Observatorio Astron\'omico Nacional on the Sierra San Pedro
M\'artir (OAN-SPM), Baja California, Mexico.
E.B. acknowledges support from DGAPA-UNAM grant IN113320.
K.M. acknowledges support from JSPS KAKENHI grant number 19K03930.
ACG's work is partially supported by the CAS ``Light of West China" Program (No. 2021-XBQNXZ-005) and the Xinjiang Tianshan Talents Program. 
G.D., O.V., M.D.J. and M.S. acknowledge support by the Astronomical
Station Vidojevica and the Ministry of Science, Technological
Development and Innovation of the Republic of Serbia (MSTDIRS) through
contract no. 451-03-66/2024-03/200002 made with Astronomical Observatory
(Belgrade), by the EC through project BELISSIMA (call FP7-REGPOT-2010-5, No. 256772), the observing and financial grant support from the Institute of Astronomy
and Rozhen NAO BAS through the bilateral SANU-BAN joint research project
"GAIA astrometry and fast variable astronomical objects", and support by
the SANU project F-187.
Y.Y.K. was supported by the M2FINDERS project which has received funding from the European Research Council (ERC) under the European Union’s Horizon2020 Research and Innovation Programme (grant agreement No 101018682).
The Submillimeter Array is a joint project between the Smithsonian Astrophysical Observatory and the Academia Sinica Institute of Astronomy and Astrophysics and is funded by the Smithsonian Institution and the Academia Sinica. We recognize that Maunakea is a culturally important site for the indigenous Hawaiian people; we are privileged to study the cosmos from its summit.
Based on observations obtained with the SARA Observatory 0.9 m telescope at Kitt Peak, which is owned and operated by the Southeastern Association for Research in Astronomy (saraobservatory.org). The authors are honored to be permitted to conduct astronomical research on Iolkam Du’ag (Kitt Peak), a mountain with particular significance to the Tohono O’odham Nation.

{\it Fermi}/LAT data were analysed with the Fermitools software (\url{https://fermi.gsfc.nasa.gov/ssc/data/analysis/documentation/}).
Wavelet software was provided by C.~Torrence and G.~Compo, and is available at URL: http://paos.colorado.edu/research/wavelets/.
Light curve simulations were based on the S.~D.~Connolly python version of the  algorithm in ref.~\cite{emma2013}, which is available at https://github.com/samconnolly/DELightcurveSimulation \cite{connolly2015}.

\end{acknowledgements}

   \bibliographystyle{aa} % style aa.bst
   \bibliography{long} % your references Yourfile.bib
\begin{appendix}

\section{Optical and radio datasets}

\begin{table*}
\caption{Details on the 43 optical datasets contributing to this paper.}
\label{tab:optical}
\centering
%\small
%\scalebox{0.9}{
\begin{tabular}{llrccr}
\hline\hline
Dataset                &  Country        & Diameter (cm)  & Symbol & Colour & $N$\\   % Autori inseriti
\hline
Abastumani             &     Georgia     &   70  & {\LARGE $\diamond$} & dark green & 2942\\  %2942
Abbey Ridge            &     Canada      &   35  & {\large $\rhd$}     & orange & 56\\ %56
Aoyama Gakuin          &     Japan       &   35  & {$\square$}         & cyan & 61\\ %61
ARIES                  &     India       &  104  & {$\square$}         & blue & 15\\  %15 
ARIES                  &     India       &  130  & {$\square$}         & green & 28\\  %28     
Athens$^a$             &     Greece      &   40  & {\LARGE $\diamond$} & cyan & 2268\\ %2268
Beli Brezi             &     Bulgaria    &   20  & {\LARGE $\ast$}     & blue & 148\\ %148
Belogradchik           &     Bulgaria    &   60  & {\large $+$}        & cyan & 94\\  %94  
Burke-Gaffney          &     Canada      &   61  & {\large $\rhd$}     & dark green & 345\\ %345
Calar Alto             &     Spain       &  220    & {\large $+$}      & yellow  & 2\\ %2
Catania (Arena)        &     Italy       &   20  & {$\times$}          & cyan & 147\\  %147  
Catania (GAC)          &     Italy       &   25  & {$\triangle$}       & cyan & 171\\  %171  
Catania (SLN)          &     Italy       &   91   & {$\triangle$}       & magenta & 230 \\ %230
Connecticut            &     US          &   51  & {\LARGE $\ast$}     & black & 458\\  %458    
Crimean (AP7p)         &     Crimea      &   70  &  {\LARGE $\circ$}   & magenta & 326\\  %326 
Crimean (ST-7)         &     Crimea      &   70  & {\large $+$}        & magenta & 51\\   %51
Crimean (ST-7; pol)    &     Crimea      &   70  & {$\times$}          & magenta & 1662\\ %1662
Felizzano              &     Italy       &   20  & {\LARGE $\ast$}     & magenta & 14\\ %14
GiaGa                  &     Italy       &   36  & {\LARGE $\ast$}     & black & 37\\ %37
Haleakala (LCO$^b$)        &     US          &   40  & {\large $+$}        & blue & 68\\  %68
Hans Haffner           &     Germany     &   50  & {\LARGE $\circ$}    & red & 1142\\  %1142    
Hypatia                &     Italy       &   25  & {\LARGE $\diamond$} & red & 4991\\ %4991    
Kitt Peak (SARA)       &     US          &   90  & {\LARGE $\diamond$} & violet & 63\\ %63
Kottamia               &     Egypt       &  188  &  {$\triangle$}      & black  & 23\\ %23
Lowell (LDT)           &     US          &  430  & {\LARGE $\circ$}    & magenta & 16\\  %16   
Lulin (SLT)            &     Taiwan      &   40  & {$\times$}          & blue & 1754\\  %1754   
McDonald (LCO$^b$)         &     US          &   40  & {$\triangle$}       & blue & 96\\   %96   
Montarrenti            &     Italy       &   53  & {\LARGE $\circ$}    & dark green & 120\\ %120    
Monte San Lorenzo      &     Italy       &   53  & {\LARGE $\circ$}    & green & 165\\ %165
Mt. Maidanak           &     Uzbekistan  &   60  & {\LARGE $\diamond$} & green & 1492\\  %1492 
New Mexico Skies       &     US          &   43  & {$\square$}         & green  & 1\\ %1
Osaka Kyoiku           &     Japan       &   51  & {$\square$}         & orange & 2061\\  %2061   
Perkins$^b$            &     US          &  180  & {\LARGE $\circ$}    & blue & 1101\\   %1101  
Pulkovo                &     Russia      &   65  & {$\times$}          & cyan &23   \\ %23
Roque (JKT)            &     Spain       &  100  & {\LARGE $\circ$}    & green  & 139 \\ %139
Roque (NOT; e2v)       &     Spain       &  256  & {\large $+$}        & green & 46\\ %46     
Rozhen                 &     Bulgaria    &  200  & {$\square$}         & red & 131\\  %131
Rozhen                 &     Bulgaria    & 50/70 & {\LARGE $\diamond$} & orange & 184\\  %184
SAI Crimean            &     Crimea      &   60  & {\LARGE $\circ$}    & orange  & 455\\ %455
San Pedro Martir       &     Mexico      &   84  & {\LARGE $\circ$}    & black & 539\\ %539
Seveso                 &     Italy       &   30  & {\large $+$}        & violet & 226\\  %226
Siena                  &     Italy       &   30  & {\LARGE $\diamond$} & blue & 2389\\    %2389 
Sierra Nevada          &     Spain       &   90  &  {\LARGE $\ast$}    & black  & 413\\ %413
Sirio                  &     Italy       &   25  & {\LARGE $\circ$}    & dark green\ & 2\\ %2
Skinakas               &     Greece      &  130  & {$\times$}          & black & 617\\   %617  
Skinakas (Robopol)     &     Greece      &  130  &  {\LARGE $\diamond$}  & black & 5  \\ %5
St.~Petersburg         &     Russia      &   40  & {\large $+$}        & orange & 1073\\   %1073  
Stocker                &     USA         &   61  & {\LARGE $\diamond$} &  black & 91 \\ %91
Svetloe                &     Russia      &   40  & {$\times$}          &  black  & 5 \\ %5
Teide (IAC80)          &     Spain       &   80  & {\LARGE $\ast$}     & green & 224\\  %224  
Teide (LCO$^b$)            &     Spain       &   40  & {\large $+$}        & black & 98\\    %98 
Teide (STELLA-I)       &     Spain       &  100  &  {\large $+$}       & violet  & 20\\ %20
Tijarafe               &     Spain       &   40  & {\LARGE $\ast$}     & red & 4241\\ %4241
Vidojevica$^c$         &     Serbia      &  140  & {$\square$}         & black & 257\\  %257   
Vidojevica$^c$         &     Serbia      &   60  & {$\triangle$}       & black & 129\\  %129   
West Mountain          &     US          &   91  & {$\triangle$}       & magenta & 985\\  %985  
Wild Boar              &     Italy       &   24  & {$\triangle$}      & green & 634\\   %634
\hline
\end{tabular}
%}
\tablefoot{
$^a$ University of Athens Observatory (UOAO); $^b$ Las Cumbres Observatory global telescope network; $^c$ Astronomical Station Vidojevica}
\end{table*}
\FloatBarrier

\begin{table*}
\caption{Details on the radio datasets contributing to this paper.}
\label{tab:radio}
\centering
\begin{tabular}{llccclr}
\hline\hline
Observatory    &  Country & Diameter & Frequency & Symbol & Colour & $N$ \\
               &          & (m)      & (GHz)     &        &        & \\
\hline
Cagliari$^a$    & Italy   & 64          &   24       & {$\triangle$} & orange & 2\\
Cagliari$^a$    & Italy   & 64          &   6.1      & {\large $\times$} & grey & 2\\
Medicina$^b$       & Italy   & 32          &   24       & {$\triangle$} & orange & 40\\
Medicina$^b$        & Italy   & 32          &   8.4      & {$\triangle$} & black & 53\\
%Mets\"ahovi     & Finland & 14          &   37       & {\LARGE $\circ$} & blue & \\
OVRO$^c$        & US      & 40          &   15       & {\large $\times$} & blue & 184\\
Pico Veleta     & Spain   & 30          &   230      & $\square$   & blue & 14\\
Pico Veleta     & Spain   & 30          &    86      & {\LARGE $\circ$} & black & 29\\
SAO RAS (RATAN-600) & Russia  & 576 (ring)  &   22.3   & {\large $+$} & dark green & 33\\
SAO RAS (RATAN-600) & Russia  & 576 (ring)  &   11.2   & {\large $+$} & red & 33\\
SAO RAS (RATAN-600) & Russia  & 576 (ring)  &   8.2  & {\large $+$} & green & 16\\
SAO RAS (RATAN-600) & Russia  & 576 (ring)  &   4.7  & {\large $+$} & magenta & 33\\
SAO RAS (RATAN-600) & Russia  & 576 (ring)  &   2.3  & {\large $+$} & cyan & 17\\
SAO RAS (RATAN-600) & Russia  & 576 (ring)  &   1.2  & {\large $+$} & violet & 5\\
Maunakea (SMA)$^d$  & US      & 8 x 6        &   345  & {\LARGE $\diamond$} & green & 5\\
Maunakea (SMA)$^d$  & US      & 8 x 6        &   273  & {$\triangle$} & magenta & 5\\
Maunakea (SMA)$^d$  & US      & 8 x 6        &   230  & {\LARGE $\diamond$} & orange & 42\\
Svetloe             & Russia  & 32          &   8.5  & {\LARGE $\circ$} & blue & 94\\
Svetloe             & Russia  & 32          &   4.8  & {\LARGE $\circ$} & orange & 132\\
%VLBA               & US      &  array      &   43   & {\LARGE $\diamond$} & red & \\
VLBA$^e$            & US      &  array      &   15   & {\LARGE $\circ$} & black & 42\\
\hline
\end{tabular}
\tablefoot{
$^a$~Sardinia Radio Telescope (SRT);
$^b$~See \cite{giroletti2020} for details on data acquisition and analysis;
$^c$~Owens Valley Radio Observatory;
$^d$~Submillimeter Array, 8-element radio interferometer with 6 m dishes;
$^e$~From the MOJAVE Program}
\end{table*}
\FloatBarrier

\section{Optical and radio light curves of BL Lacertae in the past}
\begin{figure}
  \resizebox{\hsize}{!}{\includegraphics{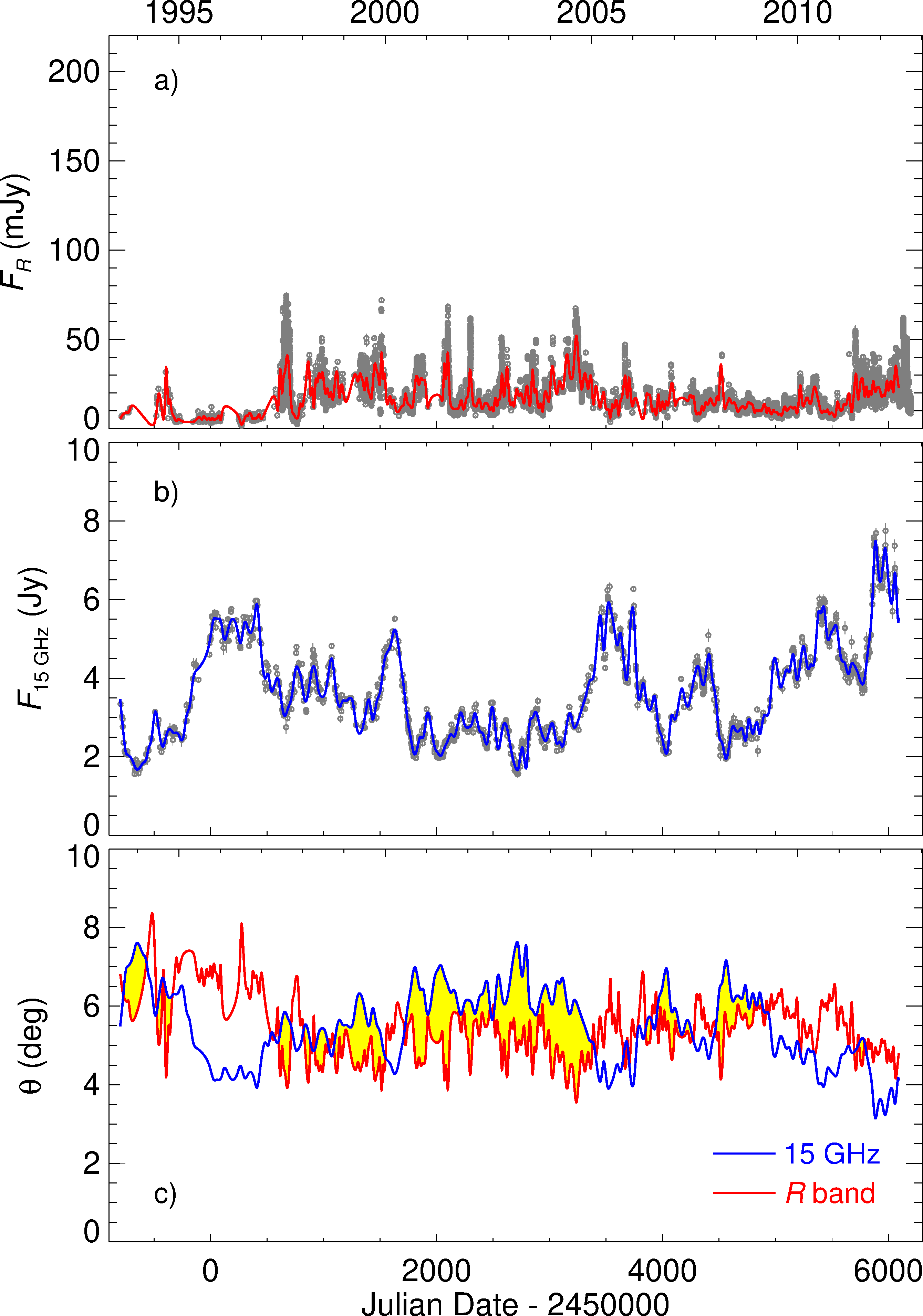}}
   \caption{Flux densities in the $R$ band (a) and at 
   %37 GHz % con Metsahovi
   15 GHz (b) in 1993--2012, and behaviour in time of the viewing angle (c) of the optical emitting region (red) and of the radio emitting region (blue). The red and blue lines in panels (a) and (b) show cubic spline interpolations on the binned optical and radio data, respectively. They represent the long-term trends of the flux.
In the bottom panels the yellow areas highlight the periods where the optical emitting region is better aligned with the line of sight than the radio zone and hence the optical radiation is more beamed than the radio emission.}
   \label{fig:radop_teta_past}%
   \end{figure}

Figure~\ref{fig:radop_teta_past} shows optical and radio data of BL Lacertae in the period 1993--2012 acquired during previous WEBT campaign on this source. 
They were used to put the data analysed in the paper into context.
The last panel displays the reconstructed twisting motion of the jet at the location of the optical and radio-emitting regions according to our model.

\FloatBarrier

\end{appendix}

\end{document}